\documentclass[%
reprint,
 superscriptaddress,
 amsmath,amssymb,
 aps,
]{revtex4-1}

\usepackage{algcompatible}
\usepackage[floatrow]{trivfloat}
\trivfloat{algorithm}

\usepackage{graphicx}
\usepackage{dcolumn}
\usepackage{bm}
\usepackage{multirow}

\usepackage{makecell}
\usepackage{listings}
\usepackage{longtable}
\usepackage{multirow}
\usepackage{siunitx}
\usepackage{xcolor}
\usepackage{floatrow}
\def\beq{\begin{eqnarray}}
\def\eeq{\end{eqnarray}}
\def\V{\mathcal{V}}

\def\P{\mathcal{P}}

\def\Nddiff{{N_{d}^{\mathrm{uniq}}}}

\def\vecD{\vec{D}}
\def\veca{\vec{\alpha}}
\def\vecb{\vec{\beta}}
\def\ia{i_\alpha}
\def\ja{j_\alpha}
\def\ib{i_\beta}

\def\vecia{\vec{i}_\alpha}

\def\veciDa{\vec{i}_{D\alpha}}
\def\veciDb{\vec{i}_{D\beta}}
\def\veciaa{\vec{i}_{\alpha\alpha}}
\def\veciba{\vec{i}_{\beta\alpha}}

\def\vecibb{\vec{i}_{\beta\beta}}

\begin{document}


\title{Fast Semistochastic Heat-Bath Configuration Interaction}

\author{Junhao Li}
  \email{jl2922@cornell.edu}
  \affiliation{Laboratory of Atomic and Solid State Physics, Cornell University, Ithaca, New York 14853, United States}
\author{Matthew Otten}
  \email{mjo98@cornell.edu}
  \affiliation{Laboratory of Atomic and Solid State Physics, Cornell University, Ithaca, New York 14853, United States}
\author{Adam A. Holmes}
  \email{aah95@cornell.edu}
  \affiliation{Laboratory of Atomic and Solid State Physics, Cornell University, Ithaca, New York 14853, United States}
  \affiliation{Department of Chemistry and Biochemistry, University of Colorado Boulder, Boulder, Colorado 80302, United States}
\author{Sandeep Sharma}
  \email{sanshar@gmail.com}
  \affiliation{Department of Chemistry and Biochemistry, University of Colorado Boulder, Boulder, Colorado 80302, United States}
\author{C. J. Umrigar}%
  \email{cyrusumrigar@cornell.edu}
  \affiliation{Laboratory of Atomic and Solid State Physics, Cornell University, Ithaca, New York 14853, United States}




\date{\today}

\begin{abstract}
This paper presents in detail our fast semistochastic heat-bath configuration interaction (SHCI) method for solving the many-body Schr\"{o}dinger equation.
We identify and eliminate computational bottlenecks in both the variational and perturbative steps of the SHCI algorithm.
We also describe the parallelization and the key data structures in our implementation, such as the distributed hash table.
The improved SHCI algorithm enables us to include in our variational wavefunction two orders of magnitude more determinants
than has been reported previously with other selected configuration interaction methods.
We use our algorithm to calculate an accurate benchmark energy for the chromium dimer with the X2C relativistic Hamiltonian in the cc-pVDZ-DK basis, correlating 28 electrons in 76 spatial orbitals.
Our largest calculation uses two billion Slater determinants in the variational space, and
semistochastically includes perturbative contributions from at least trillions of additional determinants with better than $10^{-5}$~Ha statistical uncertainty.



\end{abstract}

\maketitle

\section{Introduction}

The choice of quantum chemistry methods requires a trade-off between accuracy and efficiency.
Density functional theory
(DFT)~\cite{ParYan-BOOK-89,DreGro-BOOK-90,kohn1999nobel}
methods with approximate density functionals
are popular and efficient, but are often not sufficiently accurate.  Coupled cluster with single, double, and perturbative triple excitations CCSD(T)~\cite{raghavachari1989fifth} is very accurate for single reference systems, but
not for strongly-correlated systems, such as systems with stretched bonds.
Density matrix renormalization group (DMRG)~\cite{white1993density,white1999ab,chan2002highly,chan2011density,ShaCha-JCP-12,olivares2015ab,schollwock2005density,GuoLiCha-JCTC-18}
and full configuration interaction quantum Monte Carlo (FCIQMC)~\cite{BooThoAla-JCP-09,CleBooAla-JCP-10,PetHolChaNigUmr-PRL-12,BooGruKreAla-Nat-13,HolChaUmr-JCTC-16}
are systematically improvable but rapidly get expensive with the number of electrons and the size of the basis set.


The recently developed semistochastic heat-bath configuration interaction (SHCI)~\cite{HolTubUmr-JCTC-16,ShaHolJeaAlaUmr-JCTC-17,HolUmrSha-JCP-17,SmiMusHolSha-JCTC-17,MusSha-JCTC-17,ChiHolOttUmrShaZim-JPCA-18} is another systematically improvable method capable of providing essentially exact energies for small systems.
In common with FCIQMC, the computational cost of the method scales exponentially in the number of electrons but with a
much smaller exponent than in full configuration interation (FCI).  However, SHCI is much faster than FCIQMC.
The comparison with DMRG is more involved.
{\color{black}
While SHCI is much faster than DMRG for small moderately correlated systems, the ratio of costs changes in
DMRG's favor as the system size increases and as the correlation strength increases, because the methods
have different scaling with these parameters.
In particular SHCI scales exponentially with system size with a prefactor that is typically small, but which grows with
the strength of the correlation.
DMRG scales exponentially with the $(D-1)/D$-th power of the system size (where D is the system dimension) with a prefactor that is typically larger,
but is not very sensitive to the strength of the correlation.
}

SHCI is an example of the selected configuration interaction plus perturbation theory (SCI+PT)
methods~\cite{HurMalRan-JCP-73,BuePey-TCA-74,EvaDauMal-CP-83,CimPer-JCoP-87,Har-JCP-91,BytRue-CP-09,KelPerBarGre-JCP-14,CoeMurPat-CPL-14,Eva-JCP-14,SceAppGinCaf-JCoC-16,GarSceLooCaf-JCP-17,LooSceBloGarCafJac-JCP-18,TubLevHaiHeaWha-ARX-18},
the earliest of which being
the configuration interaction by perturbatively selecting iteratively (CIPSI) method~\cite{HurMalRan-JCP-73,EvaDauMal-CP-83} of Malrieu and collaborators.
SCI+PT methods have two stages.  In the first stage a variational wavefunction is constructed iteratively, starting from
a determinant that is expected to have a significant amplitude in the final wavefunction, e.g., the Hartree-Fock determinant.
Each iteration of the variational stage has three steps: selection of important determinants, construction of the Hamiltonian matrix, and
iterative diagonalization of the Hamiltonian matrix.
In the second stage, 2$^{\rm nd}$-order perturbation theory is used to improve upon the variational energy.

The SHCI algorithm has greatly improved the efficiency of both stages.
First, as discussed in Section~\ref{Var}, it greatly speeds up the determinant selection, and, second, as discussed in
Section~\ref{PT}, it drastically reduces the central processing unit (CPU) cost as well as the memory cost of performing the perturbation step by using a semistochastic algorithm.
These two modifications have allowed SHCI to be used for systems as large as hexatriene
in an ANO-L-pVDZ basis (32 correlated electrons in 118 orbitals) which has a Hilbert space of $10^{38}$ determinants~\cite{ChiHolOttUmrShaZim-JPCA-18}.
SHCI has also recently been extended to (a) calculate not just the ground state but also the low-lying excited states~\cite{HolUmrSha-JCP-17}, 
(b) perform self-consistent field orbital optimization in very large active spaces~\cite{SmiMusHolSha-JCTC-17},
and (c) include relativistic effects including the spin-orbit coupling using ``one-step" calculations with two-component Hamiltonians~\cite{MusSha-JCTC-17}.

Since SHCI has greatly reduced the time required to select determinants, we find, for large systems, that Hamiltonian construction is the
most time-consuming step of the variational stage.
For around $10^8$ variational determinants, it takes two orders of magnitude more time to construct the Hamiltonian matrix than to select the determinants for most molecules.
In addition, if a small stochastic error is required, the perturbative stage can be expensive,
particularly on computer systems that do not have enough memory.
Hence, in this paper, we present an improved SHCI algorithm that greatly speeds up these two steps.
For the variational stage, we introduce a fast Hamiltonian construction algorithm that allows us to use two orders of magnitude more determinants in the wavefunction.
For the perturbative stage, we introduce the 3-step batch perturbation method that further speeds up the calculation and reduces
the memory requirement.
We also describe important implementation details of the algorithm, including the key data structures and parallelization.

We organize the paper as follows:
In section~\ref{overview}, we review the SHCI method.
In section~\ref{fast}, we introduce our faster Hamiltonian construction algorithm.
In section~\ref{multi}, we introduce our 3-step batch perturbation algorithm.
In section~\ref{key}, we describe the key data structures in our implementation.
In section~\ref{para}, we describe the parallelization strategy and demonstrate its scalability.
In section~\ref{results}, we apply our improved SHCI to Cr$_2$.
Section~\ref{conclusion} concludes the paper.




\section{SHCI Review}
\label{overview}
In this section, we review the semistochastic heat-bath configuration interaction method (SHCI)~\cite{HolTubUmr-JCTC-16,ShaHolJeaAlaUmr-JCTC-17,HolUmrSha-JCP-17},
emphasizing the two important ways it differs from other SCI+PT methods.
In the following, we use $\V$ for the set of variational determinants, and $\P$ for the set of perturbative determinants, that is, the set of determinants that are connected to the variational determinants by at least one non-zero Hamiltonian matrix element but are not present in $\V$.

\subsection{Variational Stage}
\label{Var}

SHCI starts from an initial determinant
and generates the variational wave function through an iterative process.
At each iteration, the variational wavefunction, $\Psi_V$, is written as a linear combination of the determinants in the space $\V$
\begin{align}
\Psi_{V} = \sum_{D_i \in \V} c_{i} \left|D_{i}\right\rangle
\end{align}
and new determinants, ${D_a}$, from the space $\P$ that satisfy the criterion
\beq
\exists\; D_i \in \V , \mathrm{\ such\ that\ } \left|H_{a i} c_{i}\right| \ge \epsilon_{1}
\label{HCI_criterion}
\eeq
are added to the $\V$ space, where
$H_{ai}$ is the Hamiltonian matrix element between determinants $D_a$ and $D_i$, and
$\epsilon_1$ is a user-defined parameter that controls the accuracy of the variational
stage~\footnote{Since the absolute values of $c_i$ for the most important determinants tends to go down as more determinants are
included in the wavefunction, a somewhat better selection of determinants is obtained by using a larger value of
$\epsilon_1$ in the initial iterations.}.
(When $\epsilon_1=0$, the method becomes equivalent to FCI.)
After adding the new determinants to $\V$, the Hamiltonian matrix is constructed, and diagonalized using the diagonally
preconditioned Davidson method~\cite{Dav-CPC-89}, to obtain an improved estimate of the lowest eigenvalue, $E_{V}$, and eigenvector, $\Psi_V$.
This process is repeated until the change in $E_V$ falls below a certain threshold, e.g., 1~$\mu$Ha.

Other SCI methods, such as CIPSI~\cite{HurMalRan-JCP-73,EvaDauMal-CP-83} use different criteria, usually based on either the first-order perturbative
coefficient of the wavefunction,

\beq
\left|c_a^{(1)}\right|=\left|\frac{\sum_i H_{ai}c_i}{E_0-E_a}\right| > \epsilon_1
\label{eq:cipsi_ground}
\eeq
or the second-order perturbative correction to the energy.
\beq
-\Delta E_2=-\frac{\left(\sum_i H_{ai}c_i\right)^2}{E_0-E_a} > \epsilon_1.
\label{eq:cipsi_energy}
\eeq
The reason we choose instead the selection criterion in Eq.~\ref{HCI_criterion} is that it can be implemented
very efficiently without checking the vast majority of the determinants that do not meet the criterion, by taking advantage
of the fact that 
most of the Hamiltonian matrix elements correspond to double excitations, and their values do not depend
on the determinants themselves but only on the four orbitals whose occupancies change during the double excitation.
Therefore, before performing an HCI run, for each pair of spin-orbitals, the absolute values of the Hamiltonian matrix elements
obtained by doubly exciting from that pair of orbitals is computed and stored
in decreasing order by magnitude, along with the corresponding pairs of orbitals the electrons would excite to.
Then the double excitations that meet the criterion in Eq.~\ref{HCI_criterion} can be generated by
looping over all pairs of occupied orbitals in the reference determinant, and
traversing the array of sorted double-excitation matrix elements for each pair.
As soon as the cutoff is reached, the loop for that pair of occupied orbitals is exited.
Although the criterion in Eq.~\ref{HCI_criterion} does not include information from the diagonal elements,
the HCI selection criterion is not significantly different from either of the two CIPSI-like criteria because
the terms in the numerator of Eq.~\ref{eq:cipsi_ground}
span many orders of magnitude, so the sum is highly correlated with the largest-magnitude term in the sum in Eq.~\ref{eq:cipsi_ground}.
It was demonstrated in Ref.~\cite{HolTubUmr-JCTC-16} that the selected determinants give only slightly inferior convergence
to those selected using the criterion in Eq.~\ref{eq:cipsi_ground}.  This is greatly outweighed by the improved selection speed.
Moreover, one could use the HCI criterion in Eq.~\ref{HCI_criterion} with a smaller value of $\epsilon_1$ as a preselection criterion, and then select determinants
using the criterion in Eq.~\ref{eq:cipsi_energy}, thereby having the benefit of both a fast selection method and a
close to optimal choice of determinants.

\subsection{Perturbative Stage}
\label{PT}

In common with most other SCI+PT methods, the perturbative correction is
computed using Epstein-Nesbet perturbation theory~\cite{Eps-PR-26,Nes-PRS-55}.
The variational wavefunction is used to define the zeroth-order Hamiltonian, $H_0$ and the perturbation, $V$,
\begin{align}
H_0 &= \sum_{i,j \in \V} H_{ij} |D_i\rangle\langle D_j| + \sum_{a \notin \V } H_{aa} |D_a\rangle\langle D_a|. \nonumber\\
V &= H - H_0 . \label{eq:part}
\end{align}
The first-order energy correction is zero, and the second-order energy correction $\Delta E_{2}$ is
\beq
 \Delta E_{2} &=& \langle\Psi_0|V|\Psi_1\rangle
 \;=\; \sum_{a \in \P} \frac{\left(\sum_{i \in \V} H_{ai} c_i\right)^2}{E_0 - E_a},
\label{eq:PTa}
\eeq
where $E_a=H_{aa}$.

It is expensive to evaluate the expression in Eq.~\ref{eq:PTa} because the outer summation includes all determinants in the space $\P$ and their number is
${\cal O}(n^2v^2N_\V)$, where $N_\V$ is the number of variational determinants, $n$ is the number of electrons and $v$ is
the number of virtual orbitals. For the calculation on Cr$_2$, described in Section~\ref{results},
$n=28$, $v=62$ and $N_\V=2 \times 10^9$, so the number of determinants in $\P$ is huge.
The straightforward
and time-efficient approach to computing the perturbative correction requires storing
the partial sum $\sum_{i \in \V} H_{ai} c_i$ for each $a$, while
looping over all the determinants $i \in \V$. This creates a severe memory bottleneck.

Various schemes for improving the efficiency have been implemented, including only exciting from
a rediagonalized array of the largest-weight determinants~\cite{EvaDauMal-CP-83}, and its efficient approximation using
diagrammatic perturbation theory~\cite{CimPer-JCoP-87}.
However, this is both more complicated than necessary (requiring a double extrapolation with respect to the two
variational spaces to reach the Full CI limit) and is more computationally expensive than necessary since even
the largest weight determinants have many connections that make only small contributions to the energy.
The SHCI algorithm instead uses two other strategies to reduce both the computational time and the storage requirement.

First, SHCI screens the sum~\cite{HolTubUmr-JCTC-16} using a second threshold, $\epsilon_2$ (where $\epsilon_2<\epsilon_1$) as the criterion for selecting perturbative determinants $\P$,
\begin{equation}
\Delta E_{2} \left(\epsilon_{2}\right) = \sum_a \frac{\left(\sum_{D_i \in \V}^{(\epsilon_{2})}  H_{a i} c_{i}\right) ^{2}}{E_{V} - H_{a a}}
\label{eq:PTb}
\end{equation}
where $\sum^{(\epsilon_{2})}$ indicates that only terms in the sum for which $\left|H_{a i} c_{i}\right| \ge \epsilon_{2}$ are included.
Similar to the variational stage, we find the connected determinants efficiently with precomputed arrays of
double excitations sorted by the magnitude of their Hamiltonian matrix elements~\cite{HolTubUmr-JCTC-16}.
Note that the vast number of terms that do not meet this criterion are \emph{never evaluated}.

Even with this screening, the simultaneous storage of all terms indexed by $a$ in Eq.~\ref{eq:PTb} can exceed computer memory
when $\epsilon_2$ is chosen small enough to obtain essentially the exact perturbation energy.
The second innovation in the calculation of the SHCI perturbative correction is to overcome this memory bottleneck by
evaluating this perturbative correction semistochastically~\cite{ShaHolJeaAlaUmr-JCTC-17}.
The most important contributions are evaluated deterministically and the rest are sampled stochastically.
The total perturbative correction is
\beq
\Delta E_{2} \left(\epsilon_{2}\right) = \left[\Delta E_{2} ^{\mathrm{s}} \left(\epsilon_{2} \right) - \Delta E_{2} ^{\mathrm{s}} \left(\epsilon_{2} ^{\mathrm{d}}\right)\right] + \Delta E_{2} ^{\mathrm{d}} \left(\epsilon_{2} ^{\mathrm{d}}\right)
\label{eq:semistoch_PT}
\eeq
where $\Delta E_{2} ^{\mathrm{d}}$
is the deterministic perturbative correction obtained by using a larger threshold $\epsilon_2^\mathrm{d}\ge \epsilon_2$ in Eq.~\ref{eq:PTb}.
$\Delta E_{2} ^{\mathrm{s}}$
is the stochastic perturbative correction from randomly selected samples of the variational determinants, and is given by
\beq
\lefteqn{\!\!\!\!\!\!\!\!\Delta E_{2} ^{\mathrm{s}}(\epsilon_2) =
 \frac{1}{N_{d} \left(N_{d} - 1\right)} \left\langle \sum_{D_a \in \P} \left[\left(\sum_{D_i \in \V} ^{{\Nddiff}, \left(\epsilon_2\right)} \frac{w_{i} c_{i} H_{a i}}{p_{i}}\right) ^{2} \right. \right.} \nonumber \\
&&\!\!\!\!\!\!\!+ \left. \left. \sum_{D_i \in \V} ^{{\Nddiff}, \left(\epsilon_2\right)} \left(\frac{w_{i} \left(N_{d} - 1\right)}{p_{i}} - \frac{w_{i} ^{2}}{p_{i} ^{2}}\right) c_{i} ^{2} H_{a i} ^{2}\right] \frac{1}{E_{0} - E_{a}} \right\rangle
\label{eq:stoch_PT}
\eeq
where $N_d$ is the number of variational determinants per sample and
$\Nddiff$ is the number different determinants in a sample.
$p_i$ and $w_i$ are the probability of selecting determinant $D_i$ and the number of copies of that determinant in a
sample, respectively.
The $N_d$ determinants are sampled from the discrete probability distribution
\beq
p_{i} &=& {\left|c_{i}\right| \over \sum_j^{N_\V} \left|c_{j}\right|},
\label{sampling_prob}
\eeq
using the Alias method~\cite{walker1977efficient,kronmal1979alias}, which allows samples to be drawn
in ${\cal O}(1)$ time. (The more commonly used heatbath method requires ${\cal O}(\log(n))$ time to do a binary search of an array of cumulative probabilities.)
$\Delta E_2^{\mathrm{s}}[\epsilon_2]$ and $\Delta E_2^{\mathrm{s}}[\epsilon_2^{\mathrm{d}}]$ are calculated using the same set of samples,
and thus there is significant cancellation of stochastic error.
Furthermore, because these two energies are calculated simultaneously, the additional cost of performing this
calculation, compared to a purely stochastic summation, is very small.
Clearly, in the limit that $\epsilon_2^{\mathrm{d}} = \epsilon_2$, the entire perturbative calculation becomes deterministic.

The perturbative stage of the SHCI algorithm has the interesting feature that it achieves super-linear speedup with
the number of computer nodes used.  There are two reasons for this, both having to do with the increase in the total computer memory.
First, a larger fraction of the perturbative energy can be computed deterministically, using a smaller value of $\epsilon_{2} ^{\mathrm{d}}$ in Eq.~\ref{eq:semistoch_PT}.
Second, a larger value of $N_d$ in Eq.~\ref{eq:stoch_PT} can be used.
For a given total number of samples, the statistical error is smaller for a small number of large samples,
than for a large number of small samples, because the number of sampled contributions to the energy correction is a quadratic function of the number of sampled variational determinants.
{\color{black} For example, $N_s$ samples, each of size $N_d$, will have $N_s N_d^2$ contributions to the energy, whereas
$N_s/2$ samples, each of size $2N_d$, will have $2N_s N_d^2$ contributions.}
Consequently, this too contributes to a super-linear speedup.

\subsection{Other features of SHCI}
\label{other_features}
We note that although SHCI has a stochastic component, it has the advantages compared to quantum Monte Carlo algorithms that there is no sign problem,
and that each sample is independent.
Another feature of the method is that if the calculation is done for various values of the variational threshold $\epsilon_1$,
a plot of the total energy (variational plus perturbative correction) plotted versus the perturbative correction yields a smooth curve that can be used to assess the convergence and extrapolate to the Full CI limit,  $\Delta E=0$~\cite{HolUmrSha-JCP-17}.
We typically use a quadratic fit, with the points weighted by $\left(\Delta E\right)^{-2}$~\cite{ChiHolOttUmrShaZim-JPCA-18}.

As is typical in many quantum chemistry methods, we note that the convergence of both the variational energy and the total (variational plus perturbative) energy
depends on the choice of orbitals.  Natural orbitals, calculated within HCI, are typically a better choice than Hartree Fock orbitals,
and optimized orbitals~\cite{SmiMusHolSha-JCTC-17} are a yet better choice.  For systems with more than a few atoms,
split-localized optimized orbitals lead to yet better convergence~\cite{ChiHolOttUmrShaZim-JPCA-18}.

We describe, in Sections~\ref{fast} and \ref{multi}, improvements we have made to the variational and the perturbative stages of the SHCI algorithm,
which speed up the calculations by an order of magnitude or more for large systems.

\section{Fast Hamiltonian Construction}
\label{fast}

The Hamiltonian matrix is stored in upper-triangular sparse matrix form.
At each variational iteration, we have a set of old determinants, and a set of new determinants.
We have already calculated the Hamiltonian matrix for the old determinants, and need to calculate the old-new and the new-new
matrix elements.

The SHCI algorithm greatly speeds up the step of finding the important determinants and one can very quickly generate
hundreds of millions or more.  With this many determinants, the construction of the Hamiltonian matrix is expensive.
Most of the matrix elements are zero, but finding the non-zero Hamiltonian elements quickly is challenging because
the determinants in the variational wavefunction do not exhibit any pattern.
(Efficient construction of the Hamiltonian matrix of the same size in FCI is much more straightforward than in SCI.)
There are two straightforward ``brute force" approaches to building the Hamiltonian matrix:
a) looping over all pairs of determinants to find those pairs that are related by single or double excitations, and, b)
generating all connections of each determinant in $\V$ and searching for the connections in the sorted array of variational determinants.
When the number of determinants is not very large, the former is more efficient.
Both of these are much too expensive for the very large number of variational determinants that we use.

\begin{table}[h]
\caption{The notation for the data structures in our current algorithm for efficiently constructing the Hamiltonian matrix.
Analogous data structures with the alpha and beta roles reversed are also used.
The text gives details of how they are constructed efficiently.}
\begin{tabular}{ll}
\hline
\hline
Notation& \multicolumn{1}{c}{Description}\\
\hline
$\vecD$ & The array of determinants in $\V$, in the order\\
& they were generated. \\
&\\
$\veca$ & The array of all alpha strings, without repeats\\
& that are present in at least one determinant in\\
& $\V$, in the order they were generated. \\
&\\
$\ia(\alpha)$ & Hash map that takes an alpha string, $\alpha$, \\
& and returns its index, $\ia$, in $\veca$. \\
&\\
$\veciDa(\ia)$ & The array of determinant indices in $\vecD$ such\\
& that the alpha strings of those determinants \\
& have index $\ia$ in $\veca$. \\
& The elements of $\veciDa(\ia)$ are sorted either by\\
& their values, or by the indices of their beta \\
& strings in $\vecb$. (See text for details.)\\
&\\
$\veciba(\ia)$ & The array of all beta string indices in $\beta$ that\\
& appear with $\alpha_{\ia}$ in a determinant. \\
& It is sorted so that the elements of $\veciDa(\ia)$ \\
& and $\veciba(\ia)$ are always in correspondence.\\
&\\
$\vecia(\alpha^{(-1)})$ & Hash map that takes an alpha string with one \\
& less electron, $\alpha^{(-1)}$, and returns an array of  \\
& indices of alpha strings in $\veca$ that can give  \\
& $\alpha^{(-1)}$ upon removing an electron.  These are \\
& generated only for the $\alpha$'s present in the new \\
& determinants.\\
&\\
$\veciaa(j_\alpha)$ & The array of indices of alpha strings in the \\
& array $\veca$, connected by a single excitation to the \\
& $j_\alpha^{th}$ alpha string of array $\veca$, sorted in ascending \\
& order. These are generated only for the $j_\alpha$'s \\
& present in the new determinants.\\
&\\
$i_D, j_D, \cdots$ & Indices of $\vecD$.\\
&\\
$\ia, j_\alpha, \cdots$ & Indices of $\veca$.\\
&\\
$\alpha(D_i)$ & Alpha string of determinant $D_i$.\\
&\\
\hline
\end{tabular}
\label{auxiliary}
\end{table}

\begin{algorithm}
\caption{Hamiltonian matrix update for determinants connected by single or double alpha excitations.
The algorithm for single or double beta excitations is very similar.
}
\begin{algorithmic}
\FOR{$D_i$ in $\vecD$}
  \State Use hash map $\ib(\beta)$ to find $\ib$, the index of $\beta(D_i)$
  \State $s$ is the index of the first new determinant with $s > i$.
  \FOR{$j$ in $\veciDb(\ib)$}
  \IF{$j \geq s$ and $D_i, D_j$ are connected}
    \State Compute and add $H_{ij}$ to the Hamiltonian
  \ENDIF
  \ENDFOR
\ENDFOR
\end{algorithmic}
\label{same_spin}
\vskip 7mm
\end{algorithm}

\begin{algorithm}
\caption{Hamiltonian matrix update for determinants connected by an opposite-spin double excitation.
}
\begin{algorithmic}

\FOR{$D_i$ in $\vecD$}
  \State Use hash maps $\ia(\alpha)$ and $\ib(\beta)$ to find $\ia, \ib$,
  \State \hskip 4mm the indices of $\alpha(D_i)$ and $\beta(D_i)$.
  \State $s$ is the index of the first new determinant with $s>i$.
  \FOR{$k_\alpha$ in $\veciaa(\ia)$}
    \IF{number of new determinants is small}
      \FOR {$j \ge s$ in $\veciDa({k_\alpha})$ (reverse loop)}
        \IF {$\ib(\beta(D_j))\in\vecibb(\ib)$ (binary search)}
          \State Compute and add $H_{ij}$ to the Hamiltonian
        \ENDIF
      \ENDFOR
    \ELSE
      \State Find the intersection $\vec{j}_\beta$ of sorted arrays 
      \State \hskip 4mm $\veciba(k_\alpha)$ and $\vecibb(\ib)$ in ${\cal O}(n)$ time.
      \State \hskip 4mm Since $\veciba(k_\alpha)$ and $\veciDa(k_\alpha)$ are in
      \State \hskip 4mm 1-to-1 correspondence, this provides the
      \State \hskip 4mm corresponding determinants $\vec{j}_D$
      \FOR {$j$ in $\vec{j}_D$}
        \IF {$j \ge s$}
          \State Compute and add $H_{ij}$ to the Hamiltonian
        \ENDIF
      \ENDFOR
    \ENDIF
  \ENDFOR
\ENDFOR

\end{algorithmic}
\label{opposite_spin}
\end{algorithm}

The original SHCI algorithm introduced
auxiliary arrays~\cite{ShaHolJeaAlaUmr-JCTC-17} to speed up the Hamiltonian construction,
but it still spends considerable time on elements that are zero.
In our improved SHCI algorithm, we use a larger number of auxiliary arrays to further reduce the time.
All the relevant data structures are shown in Table~\ref{auxiliary}.
Some of these are appended to at each variational iteration because they contain information about all the variational determinants
currently included in the wavefunction, whereas others are constructed from scratch since part of their information content
pertains to only the new determinants.

The auxiliary arrays are constructed by looping over just the new determinants.
First, each new $\alpha$ encountered is appended to array $\veca$ and hash map $\ia(\alpha)$.
Also, each new determinant is appended to the arrays $\veciDa(\ia)$ and $\veciba(\ia)$.
In order to speed up the generation of the Hamiltonian matrix (described later) these
are sorted by $i_D$ when the number of new determinants is much smaller than the number of old determinants,
and by $\ib$
otherwise.
Then, the hash map $\vecia(\alpha^{(-1)})$ is constructed, and finally the array $\veciaa(j_\alpha)$.
The purpose of $\vecia(\alpha^{(-1)})$ is simply to speed up the construction of $\veciaa(j_\alpha)$.
Note that if two $\alpha$ strings are a single excitation apart, they will be simultaneously present under one, and only one key of the hash map $\vecia(\alpha^{(-1)})$.

Then, we update the Hamiltonian matrix using these auxiliary arrays and a loop over all the determinants.
Algorithms~\ref{same_spin} and \ref{opposite_spin} describe the algorithm using pseudocode.
The Hamiltonian matrix elements are nonzero only for determinants that are at most two excitations apart,
namely diagonal elements, same-spin single excitations, same-spin double excitations and opposite-spin double excitations.
For finding the same-spin connections, we use a method closely related
to that in Ref.~\cite{SceAppGinCaf-JCoC-16}.
Finding the opposite-spin connections is more computationally expensive and our algorithm speeds this up significantly.

\noindent \underline{\bf Same-spin excitations:} For determinants connected by single or double alpha excitations to a given determinant $D_i$, the beta strings must be the same as $\beta(D_i)$.
Hence, we simply loop over the determinants in the $\veciDb(\ib)$ array and check if the alpha strings are related by a single or a double excitation,
and if they are, we compute that Hamiltonian matrix element.
Similarly, we can find the single and double beta excitations by looping over the determinants in
the $\veciDa(\ia)$ array.

\noindent \underline{\bf Opposite-spin excitations:} For the opposite-spin double excitations, we first loop over all $k_\alpha$ in the $\veciaa(\ia)$ array,
i.e., the indices of $\veca$ connected by single excitations to $\alpha(D_i)$.
The determinants that have alpha string $k_\alpha$ are in $\veciDa(k_\alpha)$, but since only some of these
have beta strings that are single excitations of $\beta(D_i)$, we need to filter $\veciDa(k_\alpha)$
to find the connected determinants.
This is done in two different ways as described in Algorithm~\ref{opposite_spin}
depending on the number of new determinants.
When the number of new determinants is less than 20\% of the total number of determinants (e.g. in the later iterations of a given $\epsilon_1$), $\veciDa(\ia)$ and $\veciba(\ia)$ are sorted by $i_D$, otherwise, they are sorted by $\ib$.
The remaining determinants after filtering are the determinants connected to the given determinant through opposite-spin double excitations.
Each connection is visited only once during this process, which was not the case in the
original SHCI method.


\begin{figure}
  \includegraphics[width=\linewidth]{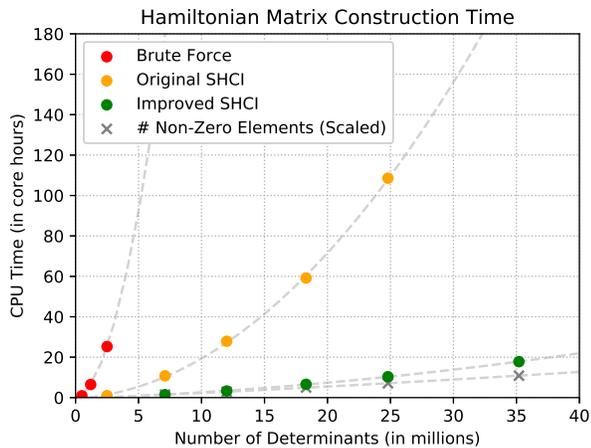}
  \caption{Hamiltonian matrix construction time for a copper atom in a cc-pVTZ basis. Hamiltonian construction is the performance bottleneck in the variational stage.
Hamiltonian construction in our improved SHCI algorithm is an order of magnitude faster than in
our original SHCI algorithm, and several orders of magnitude faster than the faster of the two brute force
approaches (loop over each pair of determinants).
Also shown is the number of nonzero elements in the Hamiltonian, scaled so that the first point coincides with
the first point of the improved SHCI CPU time.}
  \label{fig:ham}
\end{figure}

In Fig~\ref{fig:ham}, we use a copper atom with a pseudopotential~\cite{TraNee-JCP-15}~\footnote{We employ the Trail-Needs pseudopotential~\cite{TraNee-JCP-15},
but the conclusions of this paper would be the same for an all-electron calculation.}
and the cc-pVTZ basis to compare the improved SHCI algorithm to the original SHCI algorithm,
and to the brute force algorithm where we loop over each pair of determinants.
The improved algorithm
is about an order of magnitude faster than the original SHCI for medium size calculations.
For large calculations, e.g. the Cr$_2$ calculation described in Section~\ref{results}
where we use billions of variational determinants, the speedup is even greater.

\section{3-step Perturbation Energy}
\label{multi}

As described in Section~\ref{overview}, our original SHCI algorithm~\cite{ShaHolJeaAlaUmr-JCTC-17} solves the memory bottleneck problem of SCI+PT methods
by introducing a semistochastic algorithm for computing the perturbative correction to the energy.
In Section~\ref{overview} we have emphasized that the
statistical error can be dramatically reduced
by decreasing the value of $\epsilon_2^{\rm d}$ in Eq.~\ref{eq:semistoch_PT}, and by increasing the size of the stochastic samples, $N_d$ in Eq.~\ref{eq:stoch_PT}.
However, decreasing $\epsilon_2^{\rm d}$ or increasing $N_d$ can quickly lead to very large memory requirements,
making the calculations impractical even on large computers. In this situation, one is left with no choice but to run relatively
inefficient calculations with larger $\epsilon_2^{\rm d}$ and smaller $N_d$.
For example, in Table~\ref{tab:pt} rows 3 and 4 show a comparison of the total CPU time of the perturbation stage
for a copper atom in a cc-pVTZ basis on a machine with large memory versus on a machine with small memory.
When we decrease the memory by a factor of four, the total CPU time of the original SHCI algorithm increases by almost a factor of 8.


\begin{table}[h]
  \begin{tabular}{| c | c | c | c |}
  \hline
  Method & Memory & \begin{tabular}{@{}c@{}}CPU Time \\ (core hours)\end{tabular} & Error ($\mu$Ha) \\
  \hline\hline
  HCI (deterministic) & 3TB & 145.0 & 0 \\
  \hline
  \multirow{2}{*}{Original SHCI} 
  & 32GB & 116.6 & 10 \\
  \cline{2-4}
  & 128GB & 14.5 & 10 \\
  \hline
  \multirow{3}{*}{Improved SHCI}
   & 32GB & 4.2 & 9 \\
  \cline{2-4}
   & 128GB & 3.7 & 9 \\
  \cline{2-4}
   & 128GB & 5.9 & 1.8 \\
  \hline
  \end{tabular}
    \caption{
    Computational cost of perturbative correction for a copper atom in a cc-pVTZ basis.
    The variational space has 19 million determinants for $\epsilon_1=5\times10^{-5}$~Ha and the perturbative space has 35 billion determinants for $\epsilon_2=10^{-7}$~Ha.
    HCI uses the deterministic perturbation of Ref.~\onlinecite{HolTubUmr-JCTC-16}.
    SHCI uses the 2-step semistochastic perturbation algorithm of Ref.~\onlinecite{ShaHolJeaAlaUmr-JCTC-17}.
    Improved SHCI introduces the 3-step batch perturbation that significantly improves the efficiency of SHCI, especially for memory constrained cases.
    The timings for the 32GB machine are obtained by running on the same 128GB large memory machine but intentionally tuning the parameters so that the memory usage is kept below 32GB throughout the run.
    We also provide the timing to reach a 1.8~$\mu$Ha uncertainty to illustrate that our statistical error goes down much faster than $1/\sqrt{T}$ since we use smaller $\epsilon^{\rm dtm}_2$ and $\epsilon^{\rm psto}_2$ values for smaller target errors.
}
  \label{tab:pt}
\end{table}

For a given target error, assuming we have infinite computer memory, there is an optimal choice of $\epsilon_2^{\rm d}$ and $N_d$
for reaching that target error using the least computer time.
Our improved algorithm is designed to have an efficiency that depends only weakly on the available computer memory.
It is always more efficient than the original algorithm, especially when running on computers with small memory, in which case the
gain in efficiency can be orders of magnitude.
To achieve that we replace the original 2-step SHCI algorithm with a 3-step algorithm.
In each of the three steps, the perturbative determinants are divided into batches using a hash function~\cite{Jen-Hash-97, Boost-2012},
and the energy correction is computed either by adding, in succession, the contribution from each batch,
or by estimating their sum by evaluating
only a subset of these batches.
{\color{black}
The hash function maps a determinant to a 64-bit integer $h$.
A batch contains all the determinants that satisfy $h \mod n = i$, where $i$ is the batch index and $n$ is the number of batches.
We use a high-quality hash function which ensures a highly-uniform mapping,
so each batch has about the same number of determinants, i.e., the fluctuations in the number of
determinants in the various batches is the square root of the average number of determinants in each batch.
The contributions of the various batches fluctuate both because the contributions of the perturbative
determinants within a batch fluctuate and the number of perturbative determinants in a batch fluctuate.
For both contributions, the ratio of the fluctuation to the expected value is $\sim\sqrt{N}/N\to0$ for large $N$,
where $N$ is the average number of determinants in a batch.
}

In brief, our improved SHCI algorithm has the following 3 steps:
\begin{enumerate}
\item A deterministic step with cutoff $\epsilon_2^{\rm dtm} (< \epsilon_1)$, wherein
{\color{black} all the variational determinants are used}, and
all the perturbative batches are summed over.
\item A ``pseudo-stochastic" step, with cutoff $\epsilon_2^{\rm psto} (< \epsilon_2^{\rm dtm})$, wherein
{\color{black} all the variational determinants are used}, and
typically only a small fraction of the perturbative batches
need be summed over to achieve an error much smaller than the target error.
\item A stochastic step, with cutoff $\epsilon_2 (<\epsilon_2^{\rm psto}) $, wherein a few stochastic samples of variational determinants,
each consisting of $N_d$ determinants, are sampled using Eq.~\ref{sampling_prob} and only one of the
perturbative batches is randomly selected per variational sample.
\end{enumerate}
{\color{black}
The total perturbative correction is
\beq
\Delta E_{2} \left(\epsilon_{2}\right) &=& 
  \left[\Delta E_{2}^{\mathrm{sto}} \left(\epsilon_{2} \right) - \Delta E_{2}^{\mathrm{sto}} \left(\epsilon_{2}^{\rm psto}\right)\right] \nonumber \\
&+& \left[\Delta E_{2}^{\mathrm{psto}} \left(\epsilon_{2}^{\rm psto} \right) - \Delta E_{2}^{\mathrm{psto}} \left(\epsilon_{2}^{\rm dtm}\right)\right] \nonumber \\
&+& \Delta E_{2} ^{\mathrm{dtm}} \left(\epsilon_{2} ^{\mathrm{dtm}}\right)
\label{eq:semistoch_3-step_PT}
\eeq
}
The choice of these parameters depends on the system and the desired statistical error, but
reaonable choices for a target error around $10^{-5}$~Ha are
$\epsilon_2^{\rm dtm} = 2\times 10^{-6}$~Ha,
$\epsilon_2^{\rm psto} = 10^{-7}$~Ha, and,
$\epsilon_2 = \epsilon_{1}/10^6$.
{\color{black} Of course, if $\epsilon_1 \le \epsilon_2^{\rm dtm}$ the deterministic step is skipped.
}
We next describe each of the 3 steps in detail.

The first step is a deterministic step similar to the original SHCI's deterministic step,
except that when there is not enough memory to afford the
chosen $\epsilon_2^{\rm dtm}$, we divide the perturbative space into batches according to the hash value of the perturbative determinants and evaluate their contributions batch by batch.
The total deterministic correction is simply the sum of the corrections from all the batches
\beq
\Delta E_{2} \left(\epsilon_2^{\rm dtm} \right) = \sum_{B} \sum_{\substack{D_a \in \P \\ h(D_a)\in B}} \frac{\left(\sum^{(\epsilon_2^{\rm dtm})}_{D_i \in \V} H_{a i} c_{i}\right) ^{2}}{E_{V} - H_{a a}}
\eeq
where $h(D)$ is the hash function and $B$ is the hash value space for a batch.
This method solves the memory bottleneck in a different way than the original SHCI algorithm.
We could do the full calculation in this way, i.e., use a very small value for $\epsilon_2^{\rm dtm}$
{\color{black} and a large number of batches},
but it is much more efficient to only evaluate the large contributions here and leave the huge number of small contributions to the later stochastic steps.
The second step is a pseudo-stochastic step.
It is similar to the deterministic step, except for the following differences: 
a) we use an $\epsilon_2^{\rm psto}$ much smaller than $\epsilon_2^{\rm dtm}$ as the selection criterion,
b) we divide the perturbative space into as many batches as is needed in order for one batch to fit in memory,
with the constraint that there are at least 16 batches,
c) we use the corrections from the perturbative determinants in a small subset of the batches (often one is enough)
to estimate the total correction from all the perturbative determinants, as well as its standard error.
Looping over batches, for each batch, we calculate the correction from each unique perturbative determinant in that batch.
We accumulate the number of unique determinants, the sum and the sum of squares of the corrections from these determinants.
At the end of each batch iteration, we calculate the mean and standard deviation of the corrections from all the
evaluated perturbative determinants and use these to estimate the total correction from all the perturbative determinants.
Note that the standard deviation of the total correction is the standard deviation of the sum of only the unevaluated determinants.
If we process all the batches, the pseudo-stochastic step becomes deterministic and has zero standard deviation.
When the standard deviation of the total correction is smaller than 40\% of the target error, we exit the loop over batches.
However, a single batch is often sufficient to reach a statistical error below that threshold,
for the smallest $\epsilon_1$ values that we typically use.



The third step is a stochastic step that is similar to the stochastic step of the original SHCI algorithm, except that instead of keeping all the perturbative determinants
that satisfy the $\epsilon_2$ criterion we keep only one randomly selected batch out of several.
The available computer memory constrains the number of perturbative determinants, and one can obtain the same number sampling a certain
number of variational determinants and all the perturbative determinants that satisfy the $\epsilon_2$ criterion (the original SHCI algorithm),
or, by using a larger number of variational determinants and selecting just one batch of the perturbative determinants.
The latter allows us to use much larger variational samples.
Using larger variational samples is advantageous because we find that the additional fluctuations
due to sampling the perturbative determinants is much smaller than the reduction in the fluctuations due to having larger variational samples.
Typically, we use $\epsilon_2 = 10^{-6} \epsilon_{1}$.  Since the smallest $\epsilon_{1}$ that we use is typically around $10^{-5}$~Ha,
this value is in fact much smaller than is needed to ensure that the perturbative correction is fully converged.
For a statistical error of $10^{-5}$~Ha, 128 batches is usually a good choice to start with.
The size of the variational sample is chosen so that a single perturbative batch fits in the available memory.
We use a minimum of 10 samples in our stochastic step in order to get a meaningful estimate of the uncertainty.
On large memory machines, we often achieve a much smaller statistical error than the target with 10 samples.
In that case, we can decrease the size of the variational sample in later runs for similar systems.

In Table~\ref{tab:pt}, the last four rows compare the original SHCI to the improved version with 3-step batch perturbation.
In the memory-constrained case, the improved SHCI runs more than an order of magnitude faster than the original SHCI.
Even when memory is abundant, the improved SHCI is still a few times faster.

The main reasons that the improved SHCI is much faster are:
(1) It computes a larger fraction of the perturbative correction in the deterministic step.
(2) A small fraction of the batches in the pseudo-stochastic step is usually sufficient to give an accurate estimate of the total correction.
(3) It uses much larger samples of variational determinants in the stochastic step.

We now comment on a couple of aspects of our algorithm that may not be obvious:

{\color{black}1)
The value of the perturbative correction depends only on $\epsilon_2$ and not $\epsilon_2^{\rm dtm}$ and $\epsilon_2^{\rm psto}$.
The latter two quantities affect only the efficiency of the calculation.
By using batches in the stochastic step, we can use a much smaller $\epsilon_2^{\rm sto}$ and thereby
include almost the entire perturbative space.
In our calculations, we usually set $\epsilon_2^{\rm sto}=10^{-6}\epsilon_1$, which is much smaller
than is possible using our previous 2-step perturbation method, and much smaller than necessary
to keep the systematic error within the target statistical error.
}

2) In the pseudo-stochastic step, we estimate the fluctuations of the unevaluated perturbative determinants from the fluctuation of the evaluated perturbative determinants.
This relies on having a sufficiently uniform hash function.
Note that since we are using all the variational determinants in this step,
the fluctuations come just from the perturbative determinants.
In contrast, in the stochastic step, the fluctuations come both from the choice of variational determinants and the choice of batches.
In that case, one cannot simply use the standard deviation of the corrections from the evaluated perturbative determinants to estimate the standard deviation of the total correction.
So, in the stochastic step we use a minimum of 10 samples, calculate the correction from each of these samples,
and use the standard deviation of these sample corrections to estimate the standard deviation of the total correction.

{\color{black}
3) In the stochastic step, the fluctuation between batches of perturbative determinants is much smaller than the fluctuation
between samples of variational determinants.
The reason for this is that there are many more perturbative determinants in a batch (each making
only a small contribution) than there are variational determinants in a sample.
Further, the variational determinants vary greatly in importance.
This is why we use importance sampling as described by Eq.~\ref{sampling_prob} when selecting variational determinants,
and why we precede the stochastic step with the deterministic and pseudo-stochastic steps, but
even with these improvements the fluctuations from the choice of variational samples
is much larger than the fluctuation from the choice of batches.
Hence, we use only one randomly selected batch of perturbative determinants (typically out of 128 batches)
per variational sample.
}

4) The use of batches carries a small computational overhead of having to regenerate the
perturbative determinants for each batch.  Using our method, generating determinants is sufficiently fast that the
increase in computational cost would be substantial only if this is done many times.  If we employed a purely
deterministic algorithm, the number of batches would be very large, but with our 3-step semistochastic algorithm
the number of batches actually computed is sufficiently small in each of the three steps that there is never a large computational overhead.

Finally, we comment on two other algorithms that have been recently been proposed for calculating the perturbative correction.
First, another very efficient semistochastic algorithm has been proposed
by Garniron et al.~\cite{GarSceLooCaf-JCP-17}.  However, that algorithm has, for each perturbative determinant, a loop over the variational determinants to find those that are connected.  For the very large number of variational determinants that we employ
here (up to $2 \times 10^9$) this is impractical.
To avoid confusion, we should mention that the reason that their energy for Cr$_2$ is very different from ours is
that they used a nonrelativistic Hamiltonian.
Second, another algorithm that uses batches of perturbative determinants to overcome the memory bottleneck has been proposed
very recently~\cite{TubLevHaiHeaWha-ARX-18}.
It is an efficient deterministic algorithm for memory constrained environments,
but for a reasonable statistical error tolerance, e.g., $10^{-5}$~Ha,
a semistochastic approach is usually much faster, as we can see from Table~\ref{tab:pt}.
Also, in our Cr$_2$ calculation, we stochastically estimate the perturbative correction from at least trillions of perturbative determinants,
for $\epsilon_2=3\times10^{-12}$ Ha, which probably involves quadrillions of contributions
($n^2v^2N_\V = 9 \times 10^{15}$), which is infeasible with a deterministic algorithm.

\section{Key Data Structures}
\label{key}
In this section, we discuss three key data structures used to store the determinants, the distributed Hamiltonian matrix, and,
the distributed partial sums in the perturbative stage of the calculation.

\subsection{Determinants}
\label{{sec:det}}
%
%

We use two different representations of determinants.
For storing and accessing determinants locally in memory, we use arrays of bit-packed 64-bit unsigned integers.
Each bit represents a spin-orbital.
The n-th orbital is represented by the (n mod 64)-th bit of
the \lstinline{(n / 64)}-th integer, where ``\lstinline{/}'' means integer (Euclidean) division and the counting starts from zero.
(n mod 64) can be implemented as \lstinline{(n & 63)}, and \lstinline{(n / 64)} can be implemented as \lstinline{(n >> 6)},
where ``\lstinline{&}'' is the bitwise ``and'' and ``\lstinline{>>}'' is the bitwise right shift.
Both operations cost only one clock cycle on modern CPUs.

For transferring the determinants to other nodes or saving them to disk, we use base-128 variable-length integers
(VarInts)~\cite{StuFer-Protobuf-12} to compress the 64-bit integers.
VarInts take only a few bit operations to compute and reduce the memory footprint by up to 87.5\% for small integers, which reduces the network traffic and the size of the wavefunction files considerably, especially for large basis sets.

\subsection{Hamiltonian Matrix}

We store only the upper triangle of the Hamiltonian matrix.
The rows are distributed to each node in a round-robin fashion: the first row goes to the first node, the second row goes to the second node,
and when we reach the end of the node array, we loop back and start from the first node again.
Each row is a sparse vector, represented by two arrays, one stores the indices of the nonzero elements and the other stores the values.

During the matrix-vector multiplication, each node will apply its own portion of the Hamiltonian to the vector to get a partial resulting vector.
The partial results are then merged together using a binomial tree reduction.
The work on each node is distributed to the cores with dynamic load balancing.
To save space, we store only one copy of the partial resulting vector on each node and each thread updates that vector with hardware atomic operations.
In addition, we cache the diagonal of the matrix on each node to speed up the Davidson diagonalization~\cite{Dav-CPC-89}.

\subsection{Partial Sums}
%
%

In the perturbative stage, we loop over the variational determinants $\{D_i\}$ to compute the partial sum $\sum_i H_{ai}c_{i}$
for each perturbative determinant $D_a$.
The map from $D_a$ to $\sum_i H_{ai}c_{i}$, is stored in a distributed hash table~\cite{DHT}.
This choice is dictated by the enormous number of perturbative determinants we employ.
The time complexity of inserting one element into the hash table is ${\cal O}(1)$, while for a sorted array it is ${\cal O}(\log(n))$.
For large calculations, the prefactor from using hash tables is small compared to the $\log(n)$ cost from using a sorted array.

{\color{black}
The distributed hash table is based on lock-free~\cite{LockFree} open-addressing~\cite{OpenAddressing} linear-probing~\cite{LinearProbing} concurrent hash tables~\cite{chen2018concurrent} specifically designed
for intensive commutative insertion and update operations.
}
The linear-probing technique for conflict resolution has better efficiency than separate chaining during parallel insertion,
and the lock-free implementation allows all the threads to almost always operate at their full speed.

On each node, we have $n$ of these concurrent hash tables, where $n$ is the number of nodes.
One of them stores the entries belonging to that node, and the other $(n-1)$ tables store the entries belonging to other nodes pending synchronization.
Each concurrent hash table is implemented as lots of segments (at least four times the number of hardware threads) and each segment can be modified by only one thread at a time.
When a thread wants to insert or update a (key, value) pair, it first checks whether the segment that the key belongs to is being used by other threads.
If the segment is being used, the thread will insert or update the entry to a thread-local hash table, which will be merged to the main table later periodically.
We can do this because the insertion and the update operations of the partial sums are commutative.
Hence, each insertion and update is guaranteed to finish within ${\cal O}(1)$ time without getting blocked, even for perturbative determinants with lots of connections to the reference determinants.
The inter-node synchronization runs periodically so that most of the perturbative determinants will have only one copy during the entire run on the entire cluster, except for those with lots of connections to the reference determinants.

\section{Parallelization}
\label{para}

All the critical parts of SHCI are parallelized with MPI+OpenMP.
This section describes the parallelization and the scalability of each part.

When finding the connected determinants, performing the matrix-vector multiplication during the diagonalization,
and constructing the Hamiltonian matrix from the auxiliary arrays, we use the round-robin scheme to distribute the load across the nodes
and use dynamic load balancing for all the cores on the same node.


We parallelize the construction of the $\alpha$-singles, $\veciaa(j_\alpha)$, and $\beta$-singles, $\vecibb(j_\beta)$,
arrays on each node, which is the most time-consuming part of constructing the auxiliary arrays.
For each entry of $\veciaa$ and $\vecibb$, we initialize a lock to ensure exclusive modification.
We loop over all the $\vecia({\alpha^{(-1)}})$ arrays and for each ($\ia$, $\ja$) pair (which are one excitation away) inside a particular $\vecia({\alpha^{(-1)}})$ array,
we lock and append $\ja$ to $\veciaa{(\ia)}$, and
we lock and append $\ia$ to $\veciaa{(\ja)}$.
When both $\ia$ and $\ja$ occur only in the new determinants, the smaller of the two does both appends.

\begin{figure}[h]
  \includegraphics[width=\linewidth]{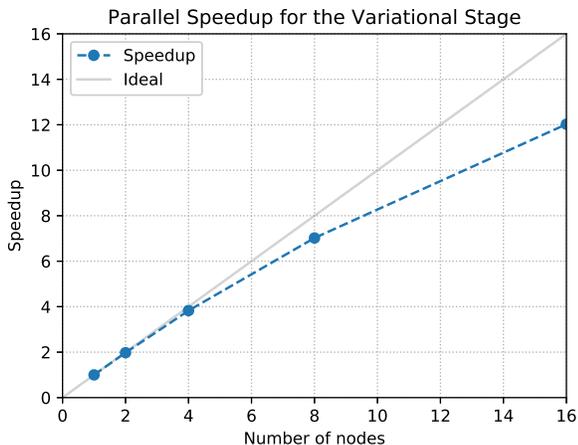}
  \caption{Parallel speedup of the variational stage for a copper atom in a cc-pVTZ basis.
There is almost perfect scaling for up to 4 nodes and 75\% parallel efficiency at 16 nodes.
}
  \label{fig:paravar}
\end{figure}

\begin{figure}[h]
  \includegraphics[width=\linewidth]{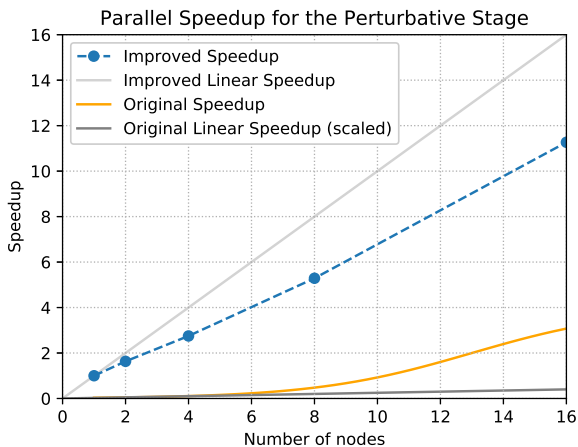}
  \caption{
Parallel speedup for improved SHCI compared to the original SHCI for the perturbative stage of the
calculation for a copper atom in a cc-pVTZ basis.
From 1 node to 4 nodes, we see a significant deviation from linear speedup due to the additional communication from shuffling the perturbative determinants across nodes.
Starting from 8 nodes, the number of shuffles approaches a constant and we can see an almost linear speedup from using more processors.
The estimate of the speedup of the original SHCI is based on the assumption that the total memory of 10 nodes is enough to support
the optimal choice of $\epsilon_2^{\rm dtm}$ and $N_d$, in Eqs.~\ref{eq:semistoch_PT} and \ref{eq:stoch_PT}.
The ``Original" curves are scaled to reflect the relative speed of original SHCI algorithm to that of the improved algorithm.
}
  \label{fig:parapt}
\end{figure}

In Figs.~\ref{fig:paravar} and \ref{fig:parapt}, we demonstrate the parallel scalability of our SHCI implementation
when applied to a copper atom in a cc-pVTZ basis. We use up to 16 nodes, and each node has 6 cores.

For the variational part, our implementation scales almost linearly up to 4 nodes.
At 16 nodes we have 75\% parallel efficiency.

For the perturbative stage, two major factors determine the speedup.
One is the additional communication associated with
shuffling perturbative determinants across the nodes, which increases with the number of nodes.
The other is the speedup from having more cores.
We can see from Fig.~\ref{fig:parapt} that from 1 to 4 nodes, the first factor dominates and there is significant deviation from ideal speedup.
Starting from 8 nodes, we have to shuffle almost all the perturbative determinants from the spawning node to the storage node that each determinant belongs to,
so there is little change in the first factor and the second factor starts to dominate, pushing the speedup curve upward and producing almost perfect scaling.
Note that in the original SHCI algorithm~\cite{ShaHolJeaAlaUmr-JCTC-17}, there is a superlinear speedup from using more nodes
because many stochastic samples are needed when running in a memory-constrained environment.
Here we have solved this problem with the 3-step batch perturbation, for which the number of stochastic samples
is almost always 10.  (We require a minimum of 10 samples in order to have a reasonable estimate of the stochastic error.)
Consequently, 
on memory constrained environments,
we achieve a few orders of magnitude speedup.

\section{Results}
\label{results}

In this section, we apply SHCI to the chromium dimer, which is a challenging strongly-correlated system
that has been studied using a variety of methods~\cite{Scu-JCP-91,KurYan-JCP-11,PurZhaKra-JCP-15,MaManOlsGag-JCTC-16,VanMalVer-JCTC-16,GuoWatHuSunCha-JCTC-16}.
We will publish the potential energy surface in a separate publication; here instead our goal is just
to use it as a test case for the improved SHCI method.
We use a relativistic exact two-component (X2C) Hamiltonian, the cc-pVDZ-DK basis, and we correlate the valence and the semi-core electrons.
This gives an active space of (28e, 76o) and a Hilbert space of $5\times10^{29}$ determinants, which is far beyond the reach of FCI.
We show how we obtain an accurate estimate of the FCI energy in this large active space with our improved SHCI algorithm.

We use PySCF~\cite{SunCha_etal_PySCF-ComMolSci-18} to generate the molecular orbital integrals for orbitals that minimize the HCI variational
energy for $\epsilon_1=2\times 10^{-4}$~Ha, using the method of Ref.~\cite{SmiMusHolSha-JCTC-17}.
We perform SHCI with several $\epsilon_1$ values from $5\times10^{-5}$ to $3\times10^{-6}$~Ha.
The Hamiltonian matrix is constructed only once.
We use very small values of $\epsilon_2 = 10^{-6} \epsilon_1$  to ensure that the perturbative correction is exceedingly well converged,
and choose the target error for the stochastic perturbation energy to be $10^{-5}$~Ha.

The improved SHCI is fast enough that we can use over two billion variational determinants, and stochastically include the contributions of at least trillions of perturbative determinants.
The largest variational calculation, where we iteratively find and diagonalize 2 billion determinants
for $\epsilon_1=3.0\times10^{-6}$~Ha, takes only one day on 8 nodes, each of which has 4 Intel Xeon E7-8870 v4 CPUs.
The corresponding perturbative calculation takes only 6 hours using only one of these nodes. 
During that perturbative calculation, we skip the deterministic step, perform a pseudo-stochastic step with $\epsilon_2^{\rm psto}=1\times10^{-7}$~Ha, and a stochastic step with $\epsilon_2=3\times10^{-12}$~Ha.
{\color{black}
We skip the deterministic step here because $\epsilon_1=3 \times 10^{-6}$ is already close to our
default $\epsilon_2^{\rm dtm}$ of $2 \times 10^{-6}$ so skipping this won't affect the efficiency
of subsequent steps much.
}
The pseudo-stochastic step uses 25 batches, each of which has about 8.9 billion determinants.
We evaluate only one of them, from which we obtain an estimate of the total correction for all the 25 batches (223 billion determinants) to be -0.011681(1)~Ha.
Since the estimated error is already much smaller than our target error, we skip the remaining 24 batches.
The pseudo-stochastic step takes 1.6 hours.
The stochastic step uses 128 batches and 6 million variational determinants in each sample,
which results in about 3.7 billion determinants per batch.
We use 10 samples and obtain the additional correction from $\epsilon_2=3.0\times10^{-12}$~Ha to be -0.001203(6)~Ha.
The combined uncertainty of the entire semistochastic perturbation stage is 6~$\mu$Ha.
It is hard to estimate how many determinants are stochastically included for $\epsilon_2=3\times10^{-12}$~Ha,
so we estimate a lower bound with $\epsilon_2^{\rm psto}=1.4\times10^{-8}$~Ha and obtain 1.8 trillion unique perturbative determinants.
Hence, with $\epsilon_2=3\times10^{-12}$~Ha (the value we are actually using) we stochastically estimate contributions from
at least trillions of unique perturbative determinants and obtain better than $10^{-5}$~Ha statistical uncertainty in 6 hours using only one node.

These large calculations enable us to obtain an estimate of the FCI energy with sub-millihartree uncertainty in this large active space.
Table~\ref{tab:Cr2} reports the results.

\begin{table}[h]
  \begin{tabular}{| c | c | c | c |}
  \hline
  $\epsilon_{1}$ (Ha) & $N_\V$ & $E_{\rm var}$ (Ha) & $E_{\rm total}$ (Ha) \\
  \hline\hline
  $5.0\times10^{-5}$ & 24M & -2099.863816 & -2099.909741(7) \\
  \hline
  $3.0\times10^{-5}$ & 53M & -2099.875327 & -2099.912356(7) \\
  \hline
  $2.0\times10^{-5}$ & 102M & -2099.883027 & -2099.914132(8) \\
  \hline
  $1.0\times10^{-5}$ & 309M & -2099.893761 & -2099.916595(1) \\
  \hline
  $7.0\times10^{-6}$ & 539M & -2099.898165 & -2099.917540(1) \\
  \hline
  $5.0\times10^{-6}$ & 911M & -2099.901781 & -2099.918306(3) \\
  \hline
  $3.0\times10^{-6}$ & 2.00B & -2099.906322 & -2099.919205(6) \\
  \hline
  0.0 (Extrap.) & - & \multicolumn{2}{ c |}{-2099.9224(6)} \\
  \hline
  \end{tabular}
  \caption{Results for Cr$_2$ at r=1.68\AA\ in the cc-pVDZ-DK basis.
  The active space is (28e, 76o).
  $N_\V$ is the number of variational determinants.
  $\epsilon_2 = 10^{-6} \epsilon_1$.
  We use weighted quadratic extrapolation, shown in Fig.~\ref{fig:extrapolation}, to obtain the FCI limit
  corresponding to $\Delta E=0$.}
  \label{tab:Cr2}
\end{table}

\begin{figure}
  \includegraphics[width=\linewidth]{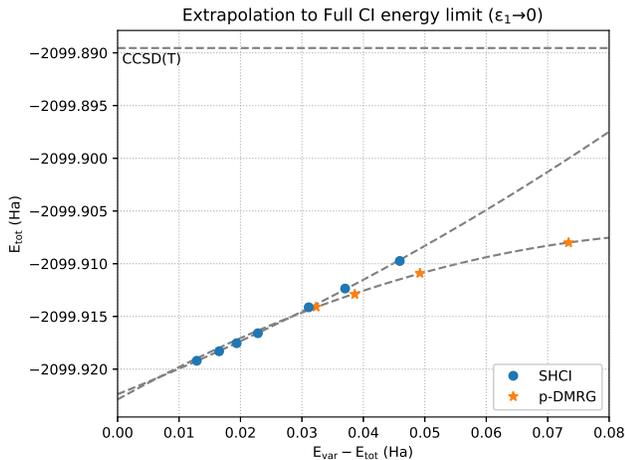}
  \caption{Weighted quadratic extrapolation of the Cr$_2$ ground state energy.
  The weight of each point is $(E_{\rm var} - E_{\rm tot})^{-2}$.
  The extrapolated energy is $-2099.9224(6)$, where the uncertainty comes from the difference between linear extrapolation and quadratic extrapolation.
  The p-DMRG extrapolation and the CCSD(T) value are also shown.
}
  \label{fig:extrapolation}
\end{figure}

We extrapolate our results using a weighted quadratic fit
and obtain for the ground state energy, $-2099.9224$~Ha as $\Delta E\to0$.
The weight of each point is $(E_{\rm var} - E_{\rm tot})^{-2}$.
Fig.~\ref{fig:extrapolation} shows the computed energies and the extrapolation.
We also perform a weighted linear fit and use the difference of the extrapolated values from the quadratic and the linear fits (0.6~mHa) as the uncertainty.
In summary, the estimated FCI energy of Cr$_2$ in the cc-pVDZ-DK basis with 28 correlated electrons and the relativistic X2C Hamiltonian
is $-2099.9224(6)$~Ha.

\begin{figure}
  \includegraphics[width=\linewidth]{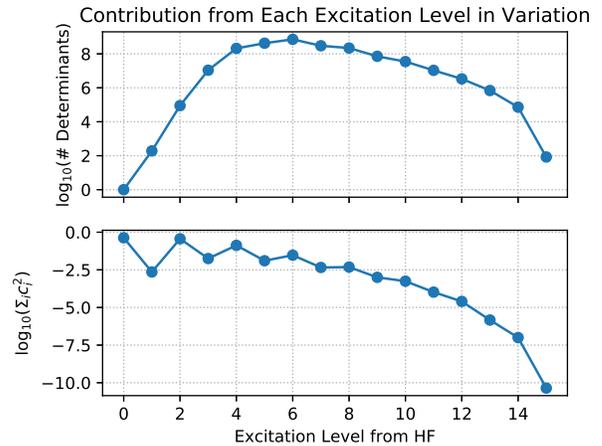}
  \caption{Contribution from each excitation level to the variational wavefunction for Cr$_2$ with $2 \times 10^9$ determinants.
  Determinants with up to 15 excitations are present in the variational wavefunction.
}
  \label{fig:excit}
\end{figure}

We compare our result with DMRG and p-DMRG, which are the only essentially exact methods that have been applied to this large active space of the chromium dimer.
The DMRG calculations use up to bond dimension $M=16000$ and obtain an extrapolated energy of $-2099.9195(27)$~Ha (default schedule) and $-2099.9192(24)$ (reverse schedule)~\cite{GuoLiCha-JCTC-18}.
These two values are similar to our most accurate data point but higher than our extrapolated result by 3~mH, which is about the estimated error of the DMRG results.
The p-DMRG calculations use up to $M=4000$ and extrapolated energy obtained from a linear fit is $-2099.9201$~Ha~\cite{GuoLiCha-JCTC-18}.
If instead, we perform a weighted quadratic fit (shown in Fig.~\ref{fig:extrapolation}), the extrapolated energy is $-2099.9225$~Ha,
in perhaps fortuitously good agreement with our result of $-2099.9224(6)$~Ha.  However, the extrapolation uncertainty is larger than the SHCI extrapolation uncertainty.
In contrast, the CCSD(T) energy is considerably higher.

One of the merits of selected-CI methods is the ability to include all excitations, regardless of excitation level.
To see the contribution from each excitation level we plot the number of selected determinants and the $\sum_i \left|c_i\right|^2$ versus excitation level in Fig.~\ref{fig:excit}.
Determinants with excitation levels up to 15 excitations are present in the variational wavefunction even though we are using optimized orbitals.
(Using Hartree-Fock orbitals, we expect that determinants with even higher excitation levels will be present.)
This implies that truncating the CI expansion at the double, triple or quadruple excitation levels (which is the most that is usually done in systematic
CI expansions), will give poor energies for such strongly correlated systems.

\section{Conclusion}
\label{conclusion}
In this paper, we introduced our fast semistochastic heat-bath configuration interaction algorithm, an efficient and essentially exact algorithm for estimating the Full-CI energy.
We introduced a new Hamiltonian generation algorithm and a 3-step batch perturbation algorithm to overcome the bottlenecks in the original SHCI algorithm.
We also presented the key data structures and parallelization strategy, which are also crucial to the performance.
These improvements allowed us to use $2 \times 10^9$ variational determinants, which is more than one order of magnitude larger than the $9\times 10^7$ determinants
used in our earlier SHCI calculation~\cite{ChiHolOttUmrShaZim-JPCA-18}, and two orders of magnitude larger than the largest variational space of $2\times 10^7$ determinants
employed to date in any other selected CI method~\cite{GarSceLooCaf-JCP-17}.

Future extensions of the method include going to yet larger variational spaces using a \emph{direct} method~\cite{Knowles1984,IvaRue-TCA-01}, wherein the Hamiltonian matrix is recalculated at each Davidson iteration and therefore need not be stored.
Although this increases the computational cost, the increase is not overwhelming because of the use of the auxiliary arrays introduced in this paper.
Other extensions include increasing the range of applicability of the method to larger systems by combining SHCI with range-separated density functional theory~\cite{Sav-INC-96},
and the use of SHCI as an impurity solver in embedding theories.
{\color{black}
Recently, a selected coupled cluster method has been developed~\cite{XuUejTen-PRL-18}.
Although the current version has
only been used with small basis sets,
it is possible that with further development this will become a highly competitive method,
especially for weakly correlated systems.
}

Possible applications of the SHCI method include providing
benchmark energies for a variety of organic molecules, as well as for transition metal atoms, dimers, and monoxides,
and calibration or training data for large scale methods,
e.g., to calibrate interatomic potentials for molecular dynamics and exchange-correlation functionals for density functional theory,
and to train machine learning based quantum chemistry solvers.
Calculations on the homogeneous electron gas are also underway.\\[1mm]

\begin{acknowledgements}
This work was supported by the AFOSR through grant FA9550-18-1-0095.
A.A.H was supported by the NSF under the grant ACI-1534965 and by the University of Colorado.
S.S. acknowledges the funding provided by the NSF through grant CHE-1800584.
The computations were performed on the Bridges computer at the Pittsburgh Supercomputing Center supported by NSF grant ACI-1445606, as part of the XSEDE program supported by NSF grant ACI-1548562.
This research also used a Director’s Discretionary allocation at the Argonne Leadership Computing Facility, which is
a DOE Office of Science User Facility supported under Contract DE-AC02-06CH11357.
\end{acknowledgements}





\bibliographystyle{apsrev4-1}

\bibliography{main,hci,umrigar,toulouse,chan,needs,zhang,chemistry}

\begin{thebibliography}{63}%
\makeatletter
\providecommand \@ifxundefined [1]{%
 \@ifx{#1\undefined}
}%
\providecommand \@ifnum [1]{%
 \ifnum #1\expandafter \@firstoftwo
 \else \expandafter \@secondoftwo
 \fi
}%
\providecommand \@ifx [1]{%
 \ifx #1\expandafter \@firstoftwo
 \else \expandafter \@secondoftwo
 \fi
}%
\providecommand \natexlab [1]{#1}%
\providecommand \enquote  [1]{``#1''}%
\providecommand \bibnamefont  [1]{#1}%
\providecommand \bibfnamefont [1]{#1}%
\providecommand \citenamefont [1]{#1}%
\providecommand \href@noop [0]{\@secondoftwo}%
\providecommand \href [0]{\begingroup \@sanitize@url \@href}%
\providecommand \@href[1]{\@@startlink{#1}\@@href}%
\providecommand \@@href[1]{\endgroup#1\@@endlink}%
\providecommand \@sanitize@url [0]{\catcode `\\12\catcode `\$12\catcode
  `\&12\catcode `\#12\catcode `\^12\catcode `\_12\catcode `\%12\relax}%
\providecommand \@@startlink[1]{}%
\providecommand \@@endlink[0]{}%
\providecommand \url  [0]{\begingroup\@sanitize@url \@url }%
\providecommand \@url [1]{\endgroup\@href {#1}{\urlprefix }}%
\providecommand \urlprefix  [0]{URL }%
\providecommand \Eprint [0]{\href }%
\providecommand \doibase [0]{http://dx.doi.org/}%
\providecommand \selectlanguage [0]{\@gobble}%
\providecommand \bibinfo  [0]{\@secondoftwo}%
\providecommand \bibfield  [0]{\@secondoftwo}%
\providecommand \translation [1]{[#1]}%
\providecommand \BibitemOpen [0]{}%
\providecommand \bibitemStop [0]{}%
\providecommand \bibitemNoStop [0]{.\EOS\space}%
\providecommand \EOS [0]{\spacefactor3000\relax}%
\providecommand \BibitemShut  [1]{\csname bibitem#1\endcsname}%
\let\auto@bib@innerbib\@empty
\bibitem [{\citenamefont {Parr}\ and\ \citenamefont
  {Yang}(1989)}]{ParYan-BOOK-89}%
  \BibitemOpen
  \bibfield  {author} {\bibinfo {author} {\bibfnamefont {R.~G.}\ \bibnamefont
  {Parr}}\ and\ \bibinfo {author} {\bibfnamefont {W.}~\bibnamefont {Yang}},\
  }\href@noop {} {\emph {\bibinfo {title} {Density-Functional Theory of Atoms
  and Molecules}}}\ (\bibinfo  {publisher} {Oxford University Press},\ \bibinfo
  {address} {New York},\ \bibinfo {year} {1989})\BibitemShut {NoStop}%
\bibitem [{\citenamefont {Dreizler}\ and\ \citenamefont
  {Gross}(1990)}]{DreGro-BOOK-90}%
  \BibitemOpen
  \bibfield  {author} {\bibinfo {author} {\bibfnamefont {R.~M.}\ \bibnamefont
  {Dreizler}}\ and\ \bibinfo {author} {\bibfnamefont {E.~K.~U.}\ \bibnamefont
  {Gross}},\ }\href@noop {} {\emph {\bibinfo {title} {Density Functional
  Theory}}}\ (\bibinfo  {publisher} {Springer-Verlag},\ \bibinfo {address}
  {Berlin},\ \bibinfo {year} {1990})\BibitemShut {NoStop}%
\bibitem [{\citenamefont {Kohn}(1999)}]{kohn1999nobel}%
  \BibitemOpen
  \bibfield  {author} {\bibinfo {author} {\bibfnamefont {W.}~\bibnamefont
  {Kohn}},\ }\href@noop {} {\bibfield  {journal} {\bibinfo  {journal} {Rev.
  Mod. Phys.}\ }\textbf {\bibinfo {volume} {71}},\ \bibinfo {pages} {1253}
  (\bibinfo {year} {1999})}\BibitemShut {NoStop}%
\bibitem [{\citenamefont {Raghavachari}\ \emph {et~al.}(1989)\citenamefont
  {Raghavachari}, \citenamefont {Trucks}, \citenamefont {Pople},\ and\
  \citenamefont {Head-Gordon}}]{raghavachari1989fifth}%
  \BibitemOpen
  \bibfield  {author} {\bibinfo {author} {\bibfnamefont {K.}~\bibnamefont
  {Raghavachari}}, \bibinfo {author} {\bibfnamefont {G.~W.}\ \bibnamefont
  {Trucks}}, \bibinfo {author} {\bibfnamefont {J.~A.}\ \bibnamefont {Pople}}, \
  and\ \bibinfo {author} {\bibfnamefont {M.}~\bibnamefont {Head-Gordon}},\
  }\href@noop {} {\bibfield  {journal} {\bibinfo  {journal} {Chemical Physics
  Letters}\ }\textbf {\bibinfo {volume} {157}},\ \bibinfo {pages} {479}
  (\bibinfo {year} {1989})}\BibitemShut {NoStop}%
\bibitem [{\citenamefont {White}(1993)}]{white1993density}%
  \BibitemOpen
  \bibfield  {author} {\bibinfo {author} {\bibfnamefont {S.~R.}\ \bibnamefont
  {White}},\ }\href@noop {} {\bibfield  {journal} {\bibinfo  {journal}
  {Physical Review B}\ }\textbf {\bibinfo {volume} {48}},\ \bibinfo {pages}
  {10345} (\bibinfo {year} {1993})}\BibitemShut {NoStop}%
\bibitem [{\citenamefont {White}\ and\ \citenamefont
  {Martin}(1999)}]{white1999ab}%
  \BibitemOpen
  \bibfield  {author} {\bibinfo {author} {\bibfnamefont {S.~R.}\ \bibnamefont
  {White}}\ and\ \bibinfo {author} {\bibfnamefont {R.~L.}\ \bibnamefont
  {Martin}},\ }\href@noop {} {\bibfield  {journal} {\bibinfo  {journal} {J.
  Chem. Phys.}\ }\textbf {\bibinfo {volume} {110}},\ \bibinfo {pages} {4127}
  (\bibinfo {year} {1999})}\BibitemShut {NoStop}%
\bibitem [{\citenamefont {Chan}\ and\ \citenamefont
  {Head-Gordon}(2002)}]{chan2002highly}%
  \BibitemOpen
  \bibfield  {author} {\bibinfo {author} {\bibfnamefont {G.~K.-L.}\
  \bibnamefont {Chan}}\ and\ \bibinfo {author} {\bibfnamefont {M.}~\bibnamefont
  {Head-Gordon}},\ }\href@noop {} {\bibfield  {journal} {\bibinfo  {journal}
  {J. Chem. Phys.}\ }\textbf {\bibinfo {volume} {116}},\ \bibinfo {pages}
  {4462} (\bibinfo {year} {2002})}\BibitemShut {NoStop}%
\bibitem [{\citenamefont {Chan}\ and\ \citenamefont
  {Sharma}(2011)}]{chan2011density}%
  \BibitemOpen
  \bibfield  {author} {\bibinfo {author} {\bibfnamefont {G.~K.-L.}\
  \bibnamefont {Chan}}\ and\ \bibinfo {author} {\bibfnamefont {S.}~\bibnamefont
  {Sharma}},\ }\href@noop {} {\bibfield  {journal} {\bibinfo  {journal} {Annu.
  Rev. Phys. Chem.}\ }\textbf {\bibinfo {volume} {62}},\ \bibinfo {pages} {465}
  (\bibinfo {year} {2011})}\BibitemShut {NoStop}%
\bibitem [{\citenamefont {Sharma}\ and\ \citenamefont
  {Chan}(2012)}]{ShaCha-JCP-12}%
  \BibitemOpen
  \bibfield  {author} {\bibinfo {author} {\bibfnamefont {S.}~\bibnamefont
  {Sharma}}\ and\ \bibinfo {author} {\bibfnamefont {G.~K.-L.}\ \bibnamefont
  {Chan}},\ }\href {\doibase {10.1063/1.3695642}} {\bibfield  {journal}
  {\bibinfo  {journal} {{J. Chem. Phys.}}\ }\textbf {\bibinfo {volume} {{136}}}
  (\bibinfo {year} {{2012}}),\ {10.1063/1.3695642}}\BibitemShut {NoStop}%
\bibitem [{\citenamefont {Olivares-Amaya}\ \emph {et~al.}(2015)\citenamefont
  {Olivares-Amaya}, \citenamefont {Hu}, \citenamefont {Nakatani}, \citenamefont
  {Sharma}, \citenamefont {Yang},\ and\ \citenamefont {Chan}}]{olivares2015ab}%
  \BibitemOpen
  \bibfield  {author} {\bibinfo {author} {\bibfnamefont {R.}~\bibnamefont
  {Olivares-Amaya}}, \bibinfo {author} {\bibfnamefont {W.}~\bibnamefont {Hu}},
  \bibinfo {author} {\bibfnamefont {N.}~\bibnamefont {Nakatani}}, \bibinfo
  {author} {\bibfnamefont {S.}~\bibnamefont {Sharma}}, \bibinfo {author}
  {\bibfnamefont {J.}~\bibnamefont {Yang}}, \ and\ \bibinfo {author}
  {\bibfnamefont {G.~K.-L.}\ \bibnamefont {Chan}},\ }\href@noop {} {\bibfield
  {journal} {\bibinfo  {journal} {J. Chem. Phys.}\ }\textbf {\bibinfo {volume}
  {142}},\ \bibinfo {pages} {034102} (\bibinfo {year} {2015})}\BibitemShut
  {NoStop}%
\bibitem [{\citenamefont {Schollw{\"o}ck}(2005)}]{schollwock2005density}%
  \BibitemOpen
  \bibfield  {author} {\bibinfo {author} {\bibfnamefont {U.}~\bibnamefont
  {Schollw{\"o}ck}},\ }\href@noop {} {\bibfield  {journal} {\bibinfo  {journal}
  {Rev. Mod. Phys.}\ }\textbf {\bibinfo {volume} {77}},\ \bibinfo {pages} {259}
  (\bibinfo {year} {2005})}\BibitemShut {NoStop}%
\bibitem [{\citenamefont {Guo}\ \emph {et~al.}(2018)\citenamefont {Guo},
  \citenamefont {Li},\ and\ \citenamefont {Chan}}]{GuoLiCha-JCTC-18}%
  \BibitemOpen
  \bibfield  {author} {\bibinfo {author} {\bibfnamefont {S.}~\bibnamefont
  {Guo}}, \bibinfo {author} {\bibfnamefont {Z.}~\bibnamefont {Li}}, \ and\
  \bibinfo {author} {\bibfnamefont {G.~K.-L.}\ \bibnamefont {Chan}},\
  }\href@noop {} {\bibfield  {journal} {\bibinfo  {journal} {J. Chem. Theory
  Comput.}\ } (\bibinfo {year} {2018})}\BibitemShut {NoStop}%
\bibitem [{\citenamefont {Booth}\ \emph {et~al.}(2009)\citenamefont {Booth},
  \citenamefont {Thom},\ and\ \citenamefont {Alavi}}]{BooThoAla-JCP-09}%
  \BibitemOpen
  \bibfield  {author} {\bibinfo {author} {\bibfnamefont {G.~H.}\ \bibnamefont
  {Booth}}, \bibinfo {author} {\bibfnamefont {A.~J.}\ \bibnamefont {Thom}}, \
  and\ \bibinfo {author} {\bibfnamefont {A.}~\bibnamefont {Alavi}},\
  }\href@noop {} {\bibfield  {journal} {\bibinfo  {journal} {J. Chem. Phys.}\
  }\textbf {\bibinfo {volume} {131}},\ \bibinfo {pages} {054106} (\bibinfo
  {year} {2009})}\BibitemShut {NoStop}%
\bibitem [{\citenamefont {Cleland}\ \emph {et~al.}(2010)\citenamefont
  {Cleland}, \citenamefont {Booth},\ and\ \citenamefont
  {Alavi}}]{CleBooAla-JCP-10}%
  \BibitemOpen
  \bibfield  {author} {\bibinfo {author} {\bibfnamefont {D.}~\bibnamefont
  {Cleland}}, \bibinfo {author} {\bibfnamefont {G.~H.}\ \bibnamefont {Booth}},
  \ and\ \bibinfo {author} {\bibfnamefont {A.}~\bibnamefont {Alavi}},\
  }\href@noop {} {\bibfield  {journal} {\bibinfo  {journal} {J. Chem. Phys.}\
  }\textbf {\bibinfo {volume} {132}},\ \bibinfo {pages} {041103} (\bibinfo
  {year} {2010})}\BibitemShut {NoStop}%
\bibitem [{\citenamefont {Petruzielo}\ \emph {et~al.}(2012)\citenamefont
  {Petruzielo}, \citenamefont {Holmes}, \citenamefont {Changlani},
  \citenamefont {Nightingale},\ and\ \citenamefont
  {Umrigar}}]{PetHolChaNigUmr-PRL-12}%
  \BibitemOpen
  \bibfield  {author} {\bibinfo {author} {\bibfnamefont {F.~R.}\ \bibnamefont
  {Petruzielo}}, \bibinfo {author} {\bibfnamefont {A.~A.}\ \bibnamefont
  {Holmes}}, \bibinfo {author} {\bibfnamefont {H.~J.}\ \bibnamefont
  {Changlani}}, \bibinfo {author} {\bibfnamefont {M.~P.}\ \bibnamefont
  {Nightingale}}, \ and\ \bibinfo {author} {\bibfnamefont {C.~J.}\ \bibnamefont
  {Umrigar}},\ }\href@noop {} {\bibfield  {journal} {\bibinfo  {journal} {Phys.
  Rev. Lett.}\ }\textbf {\bibinfo {volume} {109}},\ \bibinfo {pages} {230201}
  (\bibinfo {year} {2012})}\BibitemShut {NoStop}%
\bibitem [{\citenamefont {Booth}\ \emph {et~al.}(2013)\citenamefont {Booth},
  \citenamefont {Gr\"uneis}, \citenamefont {Kresse},\ and\ \citenamefont
  {Alavi}}]{BooGruKreAla-Nat-13}%
  \BibitemOpen
  \bibfield  {author} {\bibinfo {author} {\bibfnamefont {G.~H.}\ \bibnamefont
  {Booth}}, \bibinfo {author} {\bibfnamefont {A.}~\bibnamefont {Gr\"uneis}},
  \bibinfo {author} {\bibfnamefont {G.}~\bibnamefont {Kresse}}, \ and\ \bibinfo
  {author} {\bibfnamefont {A.}~\bibnamefont {Alavi}},\ }\href@noop {}
  {\bibfield  {journal} {\bibinfo  {journal} {Nature}\ }\textbf {\bibinfo
  {volume} {493}},\ \bibinfo {pages} {365} (\bibinfo {year}
  {2013})}\BibitemShut {NoStop}%
\bibitem [{\citenamefont {Holmes}\ \emph
  {et~al.}(2016{\natexlab{a}})\citenamefont {Holmes}, \citenamefont
  {Changlani},\ and\ \citenamefont {Umrigar}}]{HolChaUmr-JCTC-16}%
  \BibitemOpen
  \bibfield  {author} {\bibinfo {author} {\bibfnamefont {A.~A.}\ \bibnamefont
  {Holmes}}, \bibinfo {author} {\bibfnamefont {H.~J.}\ \bibnamefont
  {Changlani}}, \ and\ \bibinfo {author} {\bibfnamefont {C.~J.}\ \bibnamefont
  {Umrigar}},\ }\href@noop {} {\bibfield  {journal} {\bibinfo  {journal} {J.
  Chem. Theory Comput.}\ } (\bibinfo {year} {2016}{\natexlab{a}})}\BibitemShut
  {NoStop}%
\bibitem [{\citenamefont {Holmes}\ \emph
  {et~al.}(2016{\natexlab{b}})\citenamefont {Holmes}, \citenamefont {Tubman},\
  and\ \citenamefont {Umrigar}}]{HolTubUmr-JCTC-16}%
  \BibitemOpen
  \bibfield  {author} {\bibinfo {author} {\bibfnamefont {A.~A.}\ \bibnamefont
  {Holmes}}, \bibinfo {author} {\bibfnamefont {N.~M.}\ \bibnamefont {Tubman}},
  \ and\ \bibinfo {author} {\bibfnamefont {C.~J.}\ \bibnamefont {Umrigar}},\
  }\href@noop {} {\bibfield  {journal} {\bibinfo  {journal} {J. Chem. Theory
  Comput.}\ }\textbf {\bibinfo {volume} {{12}}},\ \bibinfo {pages} {{3674}}
  (\bibinfo {year} {{2016}}{\natexlab{b}})}\BibitemShut {NoStop}%
\bibitem [{\citenamefont {Sharma}\ \emph {et~al.}(2017)\citenamefont {Sharma},
  \citenamefont {Holmes}, \citenamefont {Jeanmairet}, \citenamefont {Alavi},\
  and\ \citenamefont {Umrigar}}]{ShaHolJeaAlaUmr-JCTC-17}%
  \BibitemOpen
  \bibfield  {author} {\bibinfo {author} {\bibfnamefont {S.}~\bibnamefont
  {Sharma}}, \bibinfo {author} {\bibfnamefont {A.~A.}\ \bibnamefont {Holmes}},
  \bibinfo {author} {\bibfnamefont {G.}~\bibnamefont {Jeanmairet}}, \bibinfo
  {author} {\bibfnamefont {A.}~\bibnamefont {Alavi}}, \ and\ \bibinfo {author}
  {\bibfnamefont {C.~J.}\ \bibnamefont {Umrigar}},\ }\href@noop {} {\bibfield
  {journal} {\bibinfo  {journal} {J. Chem. Theory Comput.}\ }\textbf {\bibinfo
  {volume} {13}},\ \bibinfo {pages} {1595} (\bibinfo {year}
  {2017})}\BibitemShut {NoStop}%
\bibitem [{\citenamefont {Holmes}\ \emph {et~al.}(2017)\citenamefont {Holmes},
  \citenamefont {Umrigar},\ and\ \citenamefont {Sharma}}]{HolUmrSha-JCP-17}%
  \BibitemOpen
  \bibfield  {author} {\bibinfo {author} {\bibfnamefont {A.~A.}\ \bibnamefont
  {Holmes}}, \bibinfo {author} {\bibfnamefont {C.~J.}\ \bibnamefont {Umrigar}},
  \ and\ \bibinfo {author} {\bibfnamefont {S.}~\bibnamefont {Sharma}},\ }\href
  {\doibase 10.1063/1.4998614} {\bibfield  {journal} {\bibinfo  {journal} {J.
  Chem. Phys.}\ }\textbf {\bibinfo {volume} {147}},\ \bibinfo {pages} {164111}
  (\bibinfo {year} {2017})}\BibitemShut {NoStop}%
\bibitem [{\citenamefont {Smith}\ \emph {et~al.}(2017)\citenamefont {Smith},
  \citenamefont {Mussard}, \citenamefont {Holmes},\ and\ \citenamefont
  {Sharma}}]{SmiMusHolSha-JCTC-17}%
  \BibitemOpen
  \bibfield  {author} {\bibinfo {author} {\bibfnamefont {J.~E.}\ \bibnamefont
  {Smith}}, \bibinfo {author} {\bibfnamefont {B.}~\bibnamefont {Mussard}},
  \bibinfo {author} {\bibfnamefont {A.~A.}\ \bibnamefont {Holmes}}, \ and\
  \bibinfo {author} {\bibfnamefont {S.}~\bibnamefont {Sharma}},\ }\href@noop {}
  {\bibfield  {journal} {\bibinfo  {journal} {J. Chem. Theory Comput.}\
  }\textbf {\bibinfo {volume} {13}},\ \bibinfo {pages} {5468} (\bibinfo {year}
  {2017})}\BibitemShut {NoStop}%
\bibitem [{\citenamefont {Mussard}\ and\ \citenamefont
  {Sharma}(2017)}]{MusSha-JCTC-17}%
  \BibitemOpen
  \bibfield  {author} {\bibinfo {author} {\bibfnamefont {B.}~\bibnamefont
  {Mussard}}\ and\ \bibinfo {author} {\bibfnamefont {S.}~\bibnamefont
  {Sharma}},\ }\href@noop {} {\bibfield  {journal} {\bibinfo  {journal} {J.
  Chem. Theory Comput.}\ }\textbf {\bibinfo {volume} {14}},\ \bibinfo {pages}
  {154} (\bibinfo {year} {2017})}\BibitemShut {NoStop}%
\bibitem [{\citenamefont {Chien}\ \emph {et~al.}(2018)\citenamefont {Chien},
  \citenamefont {Holmes}, \citenamefont {Otten}, \citenamefont {Umrigar},
  \citenamefont {Sharma},\ and\ \citenamefont
  {Zimmerman}}]{ChiHolOttUmrShaZim-JPCA-18}%
  \BibitemOpen
  \bibfield  {author} {\bibinfo {author} {\bibfnamefont {A.~D.}\ \bibnamefont
  {Chien}}, \bibinfo {author} {\bibfnamefont {A.~A.}\ \bibnamefont {Holmes}},
  \bibinfo {author} {\bibfnamefont {M.}~\bibnamefont {Otten}}, \bibinfo
  {author} {\bibfnamefont {C.~J.}\ \bibnamefont {Umrigar}}, \bibinfo {author}
  {\bibfnamefont {S.}~\bibnamefont {Sharma}}, \ and\ \bibinfo {author}
  {\bibfnamefont {P.~M.}\ \bibnamefont {Zimmerman}},\ }\href@noop {} {\bibfield
   {journal} {\bibinfo  {journal} {J. Phys. Chem. A}\ }\textbf {\bibinfo
  {volume} {122}},\ \bibinfo {pages} {2714} (\bibinfo {year}
  {2018})}\BibitemShut {NoStop}%
\bibitem [{\citenamefont {Huron}\ \emph {et~al.}(1973)\citenamefont {Huron},
  \citenamefont {Malrieu},\ and\ \citenamefont {Rancurel}}]{HurMalRan-JCP-73}%
  \BibitemOpen
  \bibfield  {author} {\bibinfo {author} {\bibfnamefont {B.}~\bibnamefont
  {Huron}}, \bibinfo {author} {\bibfnamefont {J.}~\bibnamefont {Malrieu}}, \
  and\ \bibinfo {author} {\bibfnamefont {P.}~\bibnamefont {Rancurel}},\
  }\href@noop {} {\bibfield  {journal} {\bibinfo  {journal} {J. Chem. Phys.}\
  }\textbf {\bibinfo {volume} {58}},\ \bibinfo {pages} {5745} (\bibinfo {year}
  {1973})}\BibitemShut {NoStop}%
\bibitem [{\citenamefont {Buenker}\ and\ \citenamefont
  {Peyerimhoff}(1974)}]{BuePey-TCA-74}%
  \BibitemOpen
  \bibfield  {author} {\bibinfo {author} {\bibfnamefont {R.~J.}\ \bibnamefont
  {Buenker}}\ and\ \bibinfo {author} {\bibfnamefont {S.~D.}\ \bibnamefont
  {Peyerimhoff}},\ }\href@noop {} {\bibfield  {journal} {\bibinfo  {journal}
  {Theor. Chim. Acta}\ }\textbf {\bibinfo {volume} {35}},\ \bibinfo {pages}
  {33} (\bibinfo {year} {1974})}\BibitemShut {NoStop}%
\bibitem [{\citenamefont {Evangelisti}\ \emph {et~al.}(1983)\citenamefont
  {Evangelisti}, \citenamefont {Daudey},\ and\ \citenamefont
  {Malrieu}}]{EvaDauMal-CP-83}%
  \BibitemOpen
  \bibfield  {author} {\bibinfo {author} {\bibfnamefont {S.}~\bibnamefont
  {Evangelisti}}, \bibinfo {author} {\bibfnamefont {J.-P.}\ \bibnamefont
  {Daudey}}, \ and\ \bibinfo {author} {\bibfnamefont {J.-P.}\ \bibnamefont
  {Malrieu}},\ }\href@noop {} {\bibfield  {journal} {\bibinfo  {journal} {Chem.
  Phys.}\ }\textbf {\bibinfo {volume} {75}},\ \bibinfo {pages} {91} (\bibinfo
  {year} {1983})}\BibitemShut {NoStop}%
\bibitem [{\citenamefont {Cimiraglia}\ and\ \citenamefont
  {Persico}(1987)}]{CimPer-JCoP-87}%
  \BibitemOpen
  \bibfield  {author} {\bibinfo {author} {\bibfnamefont {R.}~\bibnamefont
  {Cimiraglia}}\ and\ \bibinfo {author} {\bibfnamefont {M.}~\bibnamefont
  {Persico}},\ }\href@noop {} {\bibfield  {journal} {\bibinfo  {journal} {J.
  Comp. Chem.}\ }\textbf {\bibinfo {volume} {8}},\ \bibinfo {pages} {39}
  (\bibinfo {year} {1987})}\BibitemShut {NoStop}%
\bibitem [{\citenamefont {Harrison}(1991)}]{Har-JCP-91}%
  \BibitemOpen
  \bibfield  {author} {\bibinfo {author} {\bibfnamefont {R.~J.}\ \bibnamefont
  {Harrison}},\ }\href@noop {} {\bibfield  {journal} {\bibinfo  {journal} {J.
  Chem. Phys.}\ }\textbf {\bibinfo {volume} {94}},\ \bibinfo {pages} {5021}
  (\bibinfo {year} {1991})}\BibitemShut {NoStop}%
\bibitem [{\citenamefont {Bytautas}\ and\ \citenamefont
  {Ruedenberg}(2009)}]{BytRue-CP-09}%
  \BibitemOpen
  \bibfield  {author} {\bibinfo {author} {\bibfnamefont {L.}~\bibnamefont
  {Bytautas}}\ and\ \bibinfo {author} {\bibfnamefont {K.}~\bibnamefont
  {Ruedenberg}},\ }\href@noop {} {\bibfield  {journal} {\bibinfo  {journal}
  {Chem. Phys.}\ }\textbf {\bibinfo {volume} {356}},\ \bibinfo {pages} {64}
  (\bibinfo {year} {2009})}\BibitemShut {NoStop}%
\bibitem [{\citenamefont {Kelly}\ \emph {et~al.}(2014)\citenamefont {Kelly},
  \citenamefont {Perera}, \citenamefont {Bartlett},\ and\ \citenamefont
  {Greer}}]{KelPerBarGre-JCP-14}%
  \BibitemOpen
  \bibfield  {author} {\bibinfo {author} {\bibfnamefont {T.~P.}\ \bibnamefont
  {Kelly}}, \bibinfo {author} {\bibfnamefont {A.}~\bibnamefont {Perera}},
  \bibinfo {author} {\bibfnamefont {R.~J.}\ \bibnamefont {Bartlett}}, \ and\
  \bibinfo {author} {\bibfnamefont {J.~C.}\ \bibnamefont {Greer}},\ }\href@noop
  {} {\bibfield  {journal} {\bibinfo  {journal} {J. Chem. Phys.}\ }\textbf
  {\bibinfo {volume} {140}},\ \bibinfo {pages} {084114} (\bibinfo {year}
  {2014})}\BibitemShut {NoStop}%
\bibitem [{\citenamefont {Coe}\ \emph {et~al.}(2014)\citenamefont {Coe},
  \citenamefont {Murphy},\ and\ \citenamefont {Paterson}}]{CoeMurPat-CPL-14}%
  \BibitemOpen
  \bibfield  {author} {\bibinfo {author} {\bibfnamefont {J.}~\bibnamefont
  {Coe}}, \bibinfo {author} {\bibfnamefont {P.}~\bibnamefont {Murphy}}, \ and\
  \bibinfo {author} {\bibfnamefont {M.}~\bibnamefont {Paterson}},\ }\href@noop
  {} {\bibfield  {journal} {\bibinfo  {journal} {Chem. Phys. Lett.}\ }\textbf
  {\bibinfo {volume} {604}},\ \bibinfo {pages} {46} (\bibinfo {year}
  {2014})}\BibitemShut {NoStop}%
\bibitem [{\citenamefont {Evangelista}(2014)}]{Eva-JCP-14}%
  \BibitemOpen
  \bibfield  {author} {\bibinfo {author} {\bibfnamefont {F.~A.}\ \bibnamefont
  {Evangelista}},\ }\href@noop {} {\bibfield  {journal} {\bibinfo  {journal}
  {J. Chem. Phys.}\ }\textbf {\bibinfo {volume} {140}},\ \bibinfo {pages}
  {124114} (\bibinfo {year} {2014})}\BibitemShut {NoStop}%
\bibitem [{\citenamefont {Scemama}\ \emph {et~al.}(2016)\citenamefont
  {Scemama}, \citenamefont {Applencourt}, \citenamefont {Giner},\ and\
  \citenamefont {Caffarel}}]{SceAppGinCaf-JCoC-16}%
  \BibitemOpen
  \bibfield  {author} {\bibinfo {author} {\bibfnamefont {A.}~\bibnamefont
  {Scemama}}, \bibinfo {author} {\bibfnamefont {T.}~\bibnamefont
  {Applencourt}}, \bibinfo {author} {\bibfnamefont {E.}~\bibnamefont {Giner}},
  \ and\ \bibinfo {author} {\bibfnamefont {M.}~\bibnamefont {Caffarel}},\
  }\href@noop {} {\bibfield  {journal} {\bibinfo  {journal} {J. Comp. Chem.}\
  }\textbf {\bibinfo {volume} {37}},\ \bibinfo {pages} {1866} (\bibinfo {year}
  {2016})}\BibitemShut {NoStop}%
\bibitem [{\citenamefont {Garniron}\ \emph {et~al.}(2017)\citenamefont
  {Garniron}, \citenamefont {Scemama}, \citenamefont {Loos},\ and\
  \citenamefont {Caffarel}}]{GarSceLooCaf-JCP-17}%
  \BibitemOpen
  \bibfield  {author} {\bibinfo {author} {\bibfnamefont {Y.}~\bibnamefont
  {Garniron}}, \bibinfo {author} {\bibfnamefont {A.}~\bibnamefont {Scemama}},
  \bibinfo {author} {\bibfnamefont {P.-F.}\ \bibnamefont {Loos}}, \ and\
  \bibinfo {author} {\bibfnamefont {M.}~\bibnamefont {Caffarel}},\ }\href@noop
  {} {\bibfield  {journal} {\bibinfo  {journal} {J. Chem. Phys.}\ }\textbf
  {\bibinfo {volume} {147}},\ \bibinfo {pages} {034101} (\bibinfo {year}
  {2017})}\BibitemShut {NoStop}%
\bibitem [{\citenamefont {Loos}\ \emph {et~al.}(2018)\citenamefont {Loos},
  \citenamefont {Scemama}, \citenamefont {Blondel}, \citenamefont {Garniron},
  \citenamefont {Caffarel},\ and\ \citenamefont
  {Jacquemin}}]{LooSceBloGarCafJac-JCP-18}%
  \BibitemOpen
  \bibfield  {author} {\bibinfo {author} {\bibfnamefont {P.-F.}\ \bibnamefont
  {Loos}}, \bibinfo {author} {\bibfnamefont {A.}~\bibnamefont {Scemama}},
  \bibinfo {author} {\bibfnamefont {A.}~\bibnamefont {Blondel}}, \bibinfo
  {author} {\bibfnamefont {Y.}~\bibnamefont {Garniron}}, \bibinfo {author}
  {\bibfnamefont {M.}~\bibnamefont {Caffarel}}, \ and\ \bibinfo {author}
  {\bibfnamefont {D.}~\bibnamefont {Jacquemin}},\ }\href@noop {} {\bibfield
  {journal} {\bibinfo  {journal} {J. Chem. Theory Comput.}\ }\textbf {\bibinfo
  {volume} {14}},\ \bibinfo {pages} {4360−4379} (\bibinfo {year}
  {2018})}\BibitemShut {NoStop}%
\bibitem [{\citenamefont {Tubman}\ \emph {et~al.}()\citenamefont {Tubman},
  \citenamefont {Levine}, \citenamefont {Hait}, \citenamefont {Head-Gordon},\
  and\ \citenamefont {Whaley}}]{TubLevHaiHeaWha-ARX-18}%
  \BibitemOpen
  \bibfield  {author} {\bibinfo {author} {\bibfnamefont {N.~M.}\ \bibnamefont
  {Tubman}}, \bibinfo {author} {\bibfnamefont {D.~S.}\ \bibnamefont {Levine}},
  \bibinfo {author} {\bibfnamefont {D.}~\bibnamefont {Hait}}, \bibinfo {author}
  {\bibfnamefont {M.}~\bibnamefont {Head-Gordon}}, \ and\ \bibinfo {author}
  {\bibfnamefont {K.~B.}\ \bibnamefont {Whaley}},\ }\href@noop {} {\bibinfo
  {journal} {https://arxiv.org/pdf/1808.02049.pdf}\ }\BibitemShut {NoStop}%
\bibitem [{Note1()}]{Note1}%
  \BibitemOpen
\bibfield  {journal} {  }\bibinfo {note} {Since the absolute values of $c_i$
  for the most important determinants tends to go down as more determinants are
  included in the wavefunction, a somewhat better selection of determinants is
  obtained by using a larger value of $\epsilon _1$ in the initial
  iterations.}\BibitemShut {Stop}%
\bibitem [{\citenamefont {Davidson}(1989)}]{Dav-CPC-89}%
  \BibitemOpen
  \bibfield  {author} {\bibinfo {author} {\bibfnamefont {E.~R.}\ \bibnamefont
  {Davidson}},\ }\href@noop {} {\bibfield  {journal} {\bibinfo  {journal}
  {Computer Physics Communications}\ }\textbf {\bibinfo {volume} {53}},\
  \bibinfo {pages} {49} (\bibinfo {year} {1989})}\BibitemShut {NoStop}%
\bibitem [{\citenamefont {Epstein}(1926)}]{Eps-PR-26}%
  \BibitemOpen
  \bibfield  {author} {\bibinfo {author} {\bibfnamefont {P.~S.}\ \bibnamefont
  {Epstein}},\ }\href@noop {} {\bibfield  {journal} {\bibinfo  {journal} {Phys.
  Rev.}\ }\textbf {\bibinfo {volume} {28}},\ \bibinfo {pages} {695} (\bibinfo
  {year} {1926})}\BibitemShut {NoStop}%
\bibitem [{\citenamefont {Nesbet}(1955)}]{Nes-PRS-55}%
  \BibitemOpen
  \bibfield  {author} {\bibinfo {author} {\bibfnamefont {R.~K.}\ \bibnamefont
  {Nesbet}},\ }\href@noop {} {\bibfield  {journal} {\bibinfo  {journal} {Proc.
  R. Soc. London, Ser. A.}\ }\textbf {\bibinfo {volume} {230}},\ \bibinfo
  {pages} {312} (\bibinfo {year} {1955})}\BibitemShut {NoStop}%
\bibitem [{\citenamefont {Walker}(1977)}]{walker1977efficient}%
  \BibitemOpen
  \bibfield  {author} {\bibinfo {author} {\bibfnamefont {A.~J.}\ \bibnamefont
  {Walker}},\ }\href@noop {} {\bibfield  {journal} {\bibinfo  {journal} {ACM
  Trans. on Math. Software (TOMS)}\ }\textbf {\bibinfo {volume} {3}},\ \bibinfo
  {pages} {253} (\bibinfo {year} {1977})}\BibitemShut {NoStop}%
\bibitem [{\citenamefont {Kronmal}\ and\ \citenamefont
  {Peterson~Jr}(1979)}]{kronmal1979alias}%
  \BibitemOpen
  \bibfield  {author} {\bibinfo {author} {\bibfnamefont {R.~A.}\ \bibnamefont
  {Kronmal}}\ and\ \bibinfo {author} {\bibfnamefont {A.~V.}\ \bibnamefont
  {Peterson~Jr}},\ }\href@noop {} {\bibfield  {journal} {\bibinfo  {journal}
  {Amer. Statist.}\ }\textbf {\bibinfo {volume} {33}},\ \bibinfo {pages} {214}
  (\bibinfo {year} {1979})}\BibitemShut {NoStop}%
\bibitem [{\citenamefont {Trail}\ and\ \citenamefont
  {Needs}(2015)}]{TraNee-JCP-15}%
  \BibitemOpen
  \bibfield  {author} {\bibinfo {author} {\bibfnamefont {J.~R.}\ \bibnamefont
  {Trail}}\ and\ \bibinfo {author} {\bibfnamefont {R.~J.}\ \bibnamefont
  {Needs}},\ }\href@noop {} {\bibfield  {journal} {\bibinfo  {journal} {{J.
  Chem. Phys.}}\ }\textbf {\bibinfo {volume} {{142}}},\ \bibinfo {pages}
  {{064110}} (\bibinfo {year} {{2015}})}\BibitemShut {NoStop}%
\bibitem [{Note2()}]{Note2}%
  \BibitemOpen
  \bibinfo {note} {We employ the Trail-Needs pseudopotential~\cite
  {TraNee-JCP-15}, but the conclusions of this paper would be the same for an
  all-electron calculation.}\BibitemShut {Stop}%
\bibitem [{Jen()}]{Jen-Hash-97}%
  \BibitemOpen
  \href@noop {} {}\bibinfo {note}
  {Http://collaboration.cmc.ec.gc.ca/science/rpn/biblio/ddj/Website/articles/DDJ/1997/9709/9709n/9709n.htm}\BibitemShut
  {NoStop}%
\bibitem [{Boo()}]{Boost-2012}%
  \BibitemOpen
  \href@noop {} {}\bibinfo {note} {Boost C$++$ Libraries,
  https://www.boost.org/doc/libs/, (2012)}\BibitemShut {NoStop}%
\bibitem [{\citenamefont {Stuart}\ and\ \citenamefont
  {Fernando}(2012)}]{StuFer-Protobuf-12}%
  \BibitemOpen
  \bibfield  {author} {\bibinfo {author} {\bibfnamefont {S.}~\bibnamefont
  {Stuart}}\ and\ \bibinfo {author} {\bibfnamefont {R.}~\bibnamefont
  {Fernando}},\ }\href@noop {} {\bibfield  {journal} {\bibinfo  {journal}
  {https://tools.ietf.org/html/draft-rfernando-protocol-buffers-00}\ }
  (\bibinfo {year} {2012})}\BibitemShut {NoStop}%
\bibitem [{DHT()}]{DHT}%
  \BibitemOpen
  \href@noop {} {\enquote {\bibinfo {title} {Distributed hash table},}\
  }\bibinfo {howpublished}
  {\url{https://en.wikipedia.org/wiki/Distributed_hash_table}},\ \bibinfo
  {note} {accessed: 2018-10-12}\BibitemShut {NoStop}%
\bibitem [{Loc()}]{LockFree}%
  \BibitemOpen
  \href@noop {} {\enquote {\bibinfo {title} {Non-blocking algorithm},}\
  }\bibinfo {howpublished}
  {\url{https://en.wikipedia.org/wiki/Non-blocking_algorithm}},\ \bibinfo
  {note} {accessed: 2018-10-12}\BibitemShut {NoStop}%
\bibitem [{Ope()}]{OpenAddressing}%
  \BibitemOpen
  \href@noop {} {\enquote {\bibinfo {title} {Open addressing},}\ }\bibinfo
  {howpublished} {\url{https://en.wikipedia.org/wiki/Open_addressing}},\
  \bibinfo {note} {accessed: 2018-10-12}\BibitemShut {NoStop}%
\bibitem [{Lin()}]{LinearProbing}%
  \BibitemOpen
  \href@noop {} {\enquote {\bibinfo {title} {Linear probing},}\ }\bibinfo
  {howpublished} {\url{https://en.wikipedia.org/wiki/Linear_probing}},\
  \bibinfo {note} {accessed: 2018-10-12}\BibitemShut {NoStop}%
\bibitem [{\citenamefont {Chen}\ \emph {et~al.}(2018)\citenamefont {Chen},
  \citenamefont {He}, \citenamefont {Sun}, \citenamefont {Chen},\ and\
  \citenamefont {He}}]{chen2018concurrent}%
  \BibitemOpen
  \bibfield  {author} {\bibinfo {author} {\bibfnamefont {Z.}~\bibnamefont
  {Chen}}, \bibinfo {author} {\bibfnamefont {X.}~\bibnamefont {He}}, \bibinfo
  {author} {\bibfnamefont {J.}~\bibnamefont {Sun}}, \bibinfo {author}
  {\bibfnamefont {H.}~\bibnamefont {Chen}}, \ and\ \bibinfo {author}
  {\bibfnamefont {L.}~\bibnamefont {He}},\ }\href@noop {} {\bibfield  {journal}
  {\bibinfo  {journal} {Future Generation Computer Systems}\ }\textbf {\bibinfo
  {volume} {82}},\ \bibinfo {pages} {127} (\bibinfo {year} {2018})}\BibitemShut
  {NoStop}%
\bibitem [{\citenamefont {Scuseria}(1991)}]{Scu-JCP-91}%
  \BibitemOpen
  \bibfield  {author} {\bibinfo {author} {\bibfnamefont {G.~E.}\ \bibnamefont
  {Scuseria}},\ }\href@noop {} {\bibfield  {journal} {\bibinfo  {journal} {J.
  Chem. Phys.}\ }\textbf {\bibinfo {volume} {94}},\ \bibinfo {pages} {442}
  (\bibinfo {year} {1991})}\BibitemShut {NoStop}%
\bibitem [{\citenamefont {Kurashige}\ and\ \citenamefont
  {Yanai}(2011)}]{KurYan-JCP-11}%
  \BibitemOpen
  \bibfield  {author} {\bibinfo {author} {\bibfnamefont {Y.}~\bibnamefont
  {Kurashige}}\ and\ \bibinfo {author} {\bibfnamefont {T.}~\bibnamefont
  {Yanai}},\ }\href@noop {} {\bibfield  {journal} {\bibinfo  {journal} {J.
  Chem. Phys.}\ }\textbf {\bibinfo {volume} {135}},\ \bibinfo {pages} {094104}
  (\bibinfo {year} {2011})}\BibitemShut {NoStop}%
\bibitem [{\citenamefont {Purwanto}\ \emph {et~al.}(2015)\citenamefont
  {Purwanto}, \citenamefont {Zhang},\ and\ \citenamefont
  {Krakauer}}]{PurZhaKra-JCP-15}%
  \BibitemOpen
  \bibfield  {author} {\bibinfo {author} {\bibfnamefont {W.}~\bibnamefont
  {Purwanto}}, \bibinfo {author} {\bibfnamefont {S.}~\bibnamefont {Zhang}}, \
  and\ \bibinfo {author} {\bibfnamefont {H.}~\bibnamefont {Krakauer}},\
  }\href@noop {} {\bibfield  {journal} {\bibinfo  {journal} {J. Chem. Phys.}\
  }\textbf {\bibinfo {volume} {142}},\ \bibinfo {pages} {064302} (\bibinfo
  {year} {2015})}\BibitemShut {NoStop}%
\bibitem [{\citenamefont {Ma}\ \emph {et~al.}(2016)\citenamefont {Ma},
  \citenamefont {Manni}, \citenamefont {Olsen},\ and\ \citenamefont
  {Gagliardi}}]{MaManOlsGag-JCTC-16}%
  \BibitemOpen
  \bibfield  {author} {\bibinfo {author} {\bibfnamefont {D.}~\bibnamefont
  {Ma}}, \bibinfo {author} {\bibfnamefont {G.~L.}\ \bibnamefont {Manni}},
  \bibinfo {author} {\bibfnamefont {J.}~\bibnamefont {Olsen}}, \ and\ \bibinfo
  {author} {\bibfnamefont {L.}~\bibnamefont {Gagliardi}},\ }\href@noop {}
  {\bibfield  {journal} {\bibinfo  {journal} {J. Chem. Theory Comput.}\
  }\textbf {\bibinfo {volume} {12}},\ \bibinfo {pages} {3208} (\bibinfo {year}
  {2016})}\BibitemShut {NoStop}%
\bibitem [{\citenamefont {Vancoillie}\ \emph {et~al.}(2016)\citenamefont
  {Vancoillie}, \citenamefont {Malmqvist},\ and\ \citenamefont
  {Veryazov}}]{VanMalVer-JCTC-16}%
  \BibitemOpen
  \bibfield  {author} {\bibinfo {author} {\bibfnamefont {S.}~\bibnamefont
  {Vancoillie}}, \bibinfo {author} {\bibfnamefont {P.~A.}\ \bibnamefont
  {Malmqvist}}, \ and\ \bibinfo {author} {\bibfnamefont {V.}~\bibnamefont
  {Veryazov}},\ }\href@noop {} {\bibfield  {journal} {\bibinfo  {journal} {J.
  Chem. Theory Comput.}\ }\textbf {\bibinfo {volume} {12}},\ \bibinfo {pages}
  {1647} (\bibinfo {year} {2016})}\BibitemShut {NoStop}%
\bibitem [{\citenamefont {Guo}\ \emph {et~al.}(2016)\citenamefont {Guo},
  \citenamefont {Watson}, \citenamefont {Hu}, \citenamefont {Sun},\ and\
  \citenamefont {Chan}}]{GuoWatHuSunCha-JCTC-16}%
  \BibitemOpen
  \bibfield  {author} {\bibinfo {author} {\bibfnamefont {S.}~\bibnamefont
  {Guo}}, \bibinfo {author} {\bibfnamefont {M.~A.}\ \bibnamefont {Watson}},
  \bibinfo {author} {\bibfnamefont {W.}~\bibnamefont {Hu}}, \bibinfo {author}
  {\bibfnamefont {Q.}~\bibnamefont {Sun}}, \ and\ \bibinfo {author}
  {\bibfnamefont {G.~K.-L.}\ \bibnamefont {Chan}},\ }\href@noop {} {\bibfield
  {journal} {\bibinfo  {journal} {J. Chem. Theory Comput.}\ }\textbf {\bibinfo
  {volume} {12}},\ \bibinfo {pages} {1583} (\bibinfo {year}
  {2016})}\BibitemShut {NoStop}%
\bibitem [{\citenamefont {Sun}\ \emph {et~al.}(2018)\citenamefont {Sun},
  \citenamefont {Berkelbach}, \citenamefont {Blunt}, \citenamefont {Booth},
  \citenamefont {Guo}, \citenamefont {Li}, \citenamefont {Liu}, \citenamefont
  {McClain}, \citenamefont {Sharma}, \citenamefont {Wouters},\ and\
  \citenamefont {Chan}}]{SunCha_etal_PySCF-ComMolSci-18}%
  \BibitemOpen
  \bibfield  {author} {\bibinfo {author} {\bibfnamefont {Q.}~\bibnamefont
  {Sun}}, \bibinfo {author} {\bibfnamefont {T.~C.}\ \bibnamefont {Berkelbach}},
  \bibinfo {author} {\bibfnamefont {N.~S.}\ \bibnamefont {Blunt}}, \bibinfo
  {author} {\bibfnamefont {G.~H.}\ \bibnamefont {Booth}}, \bibinfo {author}
  {\bibfnamefont {S.}~\bibnamefont {Guo}}, \bibinfo {author} {\bibfnamefont
  {Z.}~\bibnamefont {Li}}, \bibinfo {author} {\bibfnamefont {J.}~\bibnamefont
  {Liu}}, \bibinfo {author} {\bibfnamefont {J.}~\bibnamefont {McClain}},
  \bibinfo {author} {\bibfnamefont {S.}~\bibnamefont {Sharma}}, \bibinfo
  {author} {\bibfnamefont {S.}~\bibnamefont {Wouters}}, \ and\ \bibinfo
  {author} {\bibfnamefont {G.~K.-L.}\ \bibnamefont {Chan}},\ }\href@noop {}
  {\bibfield  {journal} {\bibinfo  {journal} {WIREs Comput. Mol. Sci.}\
  }\textbf {\bibinfo {volume} {8}},\ \bibinfo {pages} {e1340} (\bibinfo {year}
  {2018})}\BibitemShut {NoStop}%
\bibitem [{\citenamefont {Knowles}\ and\ \citenamefont
  {Handy}(1984)}]{Knowles1984}%
  \BibitemOpen
  \bibfield  {author} {\bibinfo {author} {\bibfnamefont {P.}~\bibnamefont
  {Knowles}}\ and\ \bibinfo {author} {\bibfnamefont {N.}~\bibnamefont
  {Handy}},\ }\href {\doibase 10.1016/0009-2614(84)85513-X} {\bibfield
  {journal} {\bibinfo  {journal} {Chem. Phys. Lett.}\ }\textbf {\bibinfo
  {volume} {111}},\ \bibinfo {pages} {315} (\bibinfo {year}
  {1984})}\BibitemShut {NoStop}%
\bibitem [{\citenamefont {Ivanic}\ and\ \citenamefont
  {Ruedenberg}(2001)}]{IvaRue-TCA-01}%
  \BibitemOpen
  \bibfield  {author} {\bibinfo {author} {\bibfnamefont {J.}~\bibnamefont
  {Ivanic}}\ and\ \bibinfo {author} {\bibfnamefont {K.}~\bibnamefont
  {Ruedenberg}},\ }\href@noop {} {\bibfield  {journal} {\bibinfo  {journal}
  {Theor Chem Acc}\ }\textbf {\bibinfo {volume} {106}},\ \bibinfo {pages} {339}
  (\bibinfo {year} {2001})}\BibitemShut {NoStop}%
\bibitem [{\citenamefont {Savin}(1996)}]{Sav-INC-96}%
  \BibitemOpen
  \bibfield  {author} {\bibinfo {author} {\bibfnamefont {A.}~\bibnamefont
  {Savin}},\ }in\ \href@noop {} {\emph {\bibinfo {booktitle} {Recent
  Developments of Modern Density Functional Theory}}},\ \bibinfo {editor}
  {edited by\ \bibinfo {editor} {\bibfnamefont {J.~M.}\ \bibnamefont
  {Seminario}}}\ (\bibinfo  {publisher} {Elsevier},\ \bibinfo {address}
  {Amsterdam},\ \bibinfo {year} {1996})\ pp.\ \bibinfo {pages}
  {327--357}\BibitemShut {NoStop}%
\bibitem [{\citenamefont {Xu}\ \emph {et~al.}(2018)\citenamefont {Xu},
  \citenamefont {Uejima},\ and\ \citenamefont {Ten-no}}]{XuUejTen-PRL-18}%
  \BibitemOpen
  \bibfield  {author} {\bibinfo {author} {\bibfnamefont {E.}~\bibnamefont
  {Xu}}, \bibinfo {author} {\bibfnamefont {M.}~\bibnamefont {Uejima}}, \ and\
  \bibinfo {author} {\bibfnamefont {S.~L.}\ \bibnamefont {Ten-no}},\
  }\href@noop {} {\bibfield  {journal} {\bibinfo  {journal} {Phys. Rev. Lett.}\
  }\textbf {\bibinfo {volume} {121}},\ \bibinfo {pages} {113001} (\bibinfo
  {year} {2018})}\BibitemShut {NoStop}%
\end{thebibliography}%

\end{document}